\documentclass[traditabstract]{aa}

\usepackage{mathrsfs}
\usepackage[nonamebreak]{natbib}
\usepackage[stable]{footmisc}
\usepackage{txfonts}
\usepackage{xspace}
\usepackage{placeins}
\usepackage{natbib}
\bibpunct{(}{)}{;}{a}{}{,} 

\usepackage{graphicx}
\graphicspath{{Figures/}}

\usepackage{epstopdf}
\usepackage{ifthen}
\usepackage{microtype}
\usepackage[breaklinks, colorlinks, citecolor=blue]{hyperref}
\usepackage{subfigure}
\usepackage[table,usenames,dvipsnames]{xcolor}

\usepackage{amsmath,amssymb}
\usepackage[english]{babel}

\usepackage{dblfloatfix}
\usepackage{float}

\def\WMAP{{WMAP}}

\newcommand{\onesig}[1]{(68\%, \text{#1})}

\newcommand{\plik}{{\tt Plik}}

\newcommand{\mksym}[1]{\ifmmode {\rm #1}\else #1\fi}
\newcommand{\dataplus}{{+}}

\newcommand{\highL}{\mksym{highL}}

\newcommand{\TT}{\mksym{TT}}

\newcommand{\planckTTonly}{\planck\ \TT}

\newcommand{\lowTEB}{\mksym{lowP}}

\newcommand{\planckTT}{\planckTTonly\dataplus\lowTEB}

\newcommand{\As}{A_{\rm s}}

\newcommand{\ns}{n_{\rm s}}
\newcommand{\lcdm}{{$\rm{\Lambda CDM}$}}

\newcommand{\Alens}{A_{\rm L}}

\providecommand{\Planck}{\textit{Planck}}
\providecommand{\planck}{\Planck}

\providecommand{\text}[1]{\rm{#1}}

\providecommand{\muK}{\mu\rm{K}}

\providecommand{\Omb}{\Omega_{\mathrm{b}}}
\providecommand{\Omc}{\Omega_{\mathrm{c}}}

\providecommand{\CAMB}{{\tt camb}}
\providecommand{\COSMOMC}{{\tt CosmoMC}}

\providecommand{\CLASS}{{\tt class}}

\newcommand{\begm}{\begin{pmatrix}}
\newcommand{\enm}{\end{pmatrix}}

\newcommand\ba{\begin{eqnarray}}
\newcommand\ea{\end{eqnarray}}
\newcommand\bea{\begin{eqnarray}}
\newcommand\eea{\end{eqnarray}}

\newcommand\be{\begin{equation}}
\newcommand\ee{\end{equation}}

\def\pmb#1{\setbox0=\hbox{#1}%
    \kern-.025em\copy0\kern-\wd0
    \kern.05em\copy0\kern-\wd0
    \kern-.025em\raise.0433em\box0}

\def\muk{\ensuremath{(\mu{\rm K})^2}}

\def\p2Y{\;_2Y}
\def\m2Y{\;_{-2}Y}
\def\beglet{
  \addtocounter{equation}{1}%
  \setcounter{parentequation}{\value{equation}}%
  \setcounter{equation}{0}%
  \def\theequation{\arabic{parentequation}\alph{equation}}%
  \ignorespaces
}
\def\endlet{
  \setcounter{equation}{\value{parentequation}}%
  \def\theequation{\arabic{equation}}%
}
\providecommand{\beglet}{\begin{subequations}}
\providecommand{\endlet}{\end{subequations}}

\def\setsymbol#1#2{\expandafter\def\csname #1\endcsname{#2}}
\def\getsymbol#1{\csname #1\endcsname}

\def\Planck{\textit{Planck}}

\newbox\tablebox    \newdimen\tablewidth
\def\leaderfil{\leaders\hbox to 5pt{\hss.\hss}\hfil}
\def\endPlancktable{\tablewidth=\columnwidth 
    $$\hss\copy\tablebox\hss$$
    \vskip-\lastskip\vskip -2pt}
\def\endPlancktablewide{\tablewidth=\textwidth 
    $$\hss\copy\tablebox\hss$$
    \vskip-\lastskip\vskip -2pt}
\def\tablenote#1 #2\par{\begingroup \parindent=0.8em
    \abovedisplayshortskip=0pt\belowdisplayshortskip=0pt
    \noindent
    $$\hss\vbox{\hsize\tablewidth \hangindent=\parindent \hangafter=1 \noindent
    \hbox to \parindent{$^#1$\hss}\strut#2\strut\par}\hss$$
    \endgroup}
\def\doubleline{\vskip 3pt\hrule \vskip 1.5pt \hrule \vskip 5pt}

\def\L2{\ifmmode L_2\else $L_2$\fi}

\def\DeltaT{\ifmmode \Delta T\else $\Delta T$\fi}
\def\deltat{\ifmmode \Delta t\else $\Delta t$\fi}
\def\fknee{\ifmmode f_{\rm knee}\else $f_{\rm knee}$\fi}
\def\Fmax{\ifmmode F_{\rm max}\else $F_{\rm max}$\fi}
\def\solar{\ifmmode{\rm M}_{\mathord\odot}\else${\rm M}_{\mathord\odot}$\fi}
\def\Msolar{\ifmmode{\rm M}_{\mathord\odot}\else${\rm M}_{\mathord\odot}$\fi}
\def\Lsolar{\ifmmode{\rm L}_{\mathord\odot}\else${\rm L}_{\mathord\odot}$\fi}
\def\inv{\ifmmode^{-1}\else$^{-1}$\fi}
\def\mo{\ifmmode^{-1}\else$^{-1}$\fi}
\def\sup#1{\ifmmode ^{\rm #1}\else $^{\rm #1}$\fi}
\def\expo#1{\ifmmode \times 10^{#1}\else $\times 10^{#1}$\fi}
\def\,{\thinspace}
\def\lsim{\mathrel{\raise .4ex\hbox{\rlap{$<$}\lower 1.2ex\hbox{$\sim$}}}}
\def\gsim{\mathrel{\raise .4ex\hbox{\rlap{$>$}\lower 1.2ex\hbox{$\sim$}}}}

\def\simprop{\mathrel{\raise .4ex\hbox{\rlap{$\propto$}\lower 1.2ex\hbox{$\sim$}}}}
\def\deg{\ifmmode^\circ\else$^\circ$\fi}
\def\pdeg{\ifmmode $\setbox0=\hbox{$^{\circ}$}\rlap{\hskip.11\wd0 .}$^{\circ}
          \else \setbox0=\hbox{$^{\circ}$}\rlap{\hskip.11\wd0 .}$^{\circ}$\fi}
\def\arcs{\ifmmode {^{\scriptstyle\prime\prime}}
          \else $^{\scriptstyle\prime\prime}$\fi}
\def\arcm{\ifmmode {^{\scriptstyle\prime}}
          \else $^{\scriptstyle\prime}$\fi}
\newdimen\sa  \newdimen\sb
\def\parcs{\sa=.07em \sb=.03em
     \ifmmode \hbox{\rlap{.}}^{\scriptstyle\prime\kern -\sb\prime}\hbox{\kern -\sa}
     \else \rlap{.}$^{\scriptstyle\prime\kern -\sb\prime}$\kern -\sa\fi}
\def\parcm{\sa=.08em \sb=.03em
     \ifmmode \hbox{\rlap{.}\kern\sa}^{\scriptstyle\prime}\hbox{\kern-\sb}
     \else \rlap{.}\kern\sa$^{\scriptstyle\prime}$\kern-\sb\fi}
\def\ra[#1 #2 #3.#4]{#1\sup{h}#2\sup{m}#3\sup{s}\llap.#4}
\def\dec[#1 #2 #3.#4]{#1\deg#2\arcm#3\arcs\llap.#4}
\def\deco[#1 #2 #3]{#1\deg#2\arcm#3\arcs}
\def\rra[#1 #2]{#1\sup{h}#2\sup{m}}

\def\dots{\relax\ifmmode \ldots\else $\ldots$\fi}
\def\WHzsr{\ifmmode $W\,Hz\mo\,sr\mo$\else W\,Hz\mo\,sr\mo\fi}
\def\mHz{\ifmmode $\,mHz$\else \,mHz\fi}
\def\GHz{\ifmmode $\,GHz$\else \,GHz\fi}
\def\mKs{\ifmmode $\,mK\,s$^{1/2}\else \,mK\,s$^{1/2}$\fi}
\def\muKs{\ifmmode \,\mu$K\,s$^{1/2}\else \,$\mu$K\,s$^{1/2}$\fi}
\def\muKRJs{\ifmmode \,\mu$K$_{\rm RJ}$\,s$^{1/2}\else \,$\mu$K$_{\rm RJ}$\,s$^{1/2}$\fi}
\def\muKHz{\ifmmode \,\mu$K\,Hz$^{-1/2}\else \,$\mu$K\,Hz$^{-1/2}$\fi}
\def\MJysr{\ifmmode \,$MJy\,sr\mo$\else \,MJy\,sr\mo\fi}
\def\MJysrmK{\ifmmode \,$MJy\,sr\mo$\,mK$_{\rm CMB}\mo\else \,MJy\,sr\mo\,mK$_{\rm CMB}\mo$\fi}
\def\microns{\ifmmode \,\mu$m$\else \,$\mu$m\fi}

\def\muK{\ifmmode \,\mu$K$\else \,$\mu$\hbox{K}\fi}
\def\microK{\ifmmode \,\mu$K$\else \,$\mu$\hbox{K}\fi}
\def\muW{\ifmmode \,\mu$W$\else \,$\mu$\hbox{W}\fi}
\def\kms{\ifmmode $\,km\,s$^{-1}\else \,km\,s$^{-1}$\fi}
\def\kmsMpc{\ifmmode $\,\kms\,Mpc\mo$\else \,\kms\,Mpc\mo\fi}
\providecommand{\sorthelp}[1]{}

\defcitealias{planck2014-a15}{PCP15}
\newcommand{\params}{\citetalias{planck2014-a15}}

\newcommand{\pip}{\cite{planck2013-XVI}}

\defcitealias{planck2014-a13}{Like15}

\newcommand{\liket}{\citetalias{planck2014-a13}}

\usepackage{booktabs}

\newcommand{\PLA}{\href{http://www.cosmos.esa.int/web/planck/pla}{http://www.cosmos.esa.int/web/planck/pla}}
\newcommand{\wwwcamb}{\href{http://camb.info}{http://camb.info}}
\newcommand{\wwwclass}{\href{http://class-code.net}{http://class-code.net}}
\newcommand{\wwwminuit}{\href{http://seal.web.cern.ch/seal/work-packages/mathlibs/minuit/index.html}{http://seal.web.cern.ch/seal/work-packages/mathlibs/minuit/index.html}}

\newcommand{\citeg}[1]{\citep[e.g.,][]{#1}}

\renewcommand{\onesig}[1]{(\text{#1})}

\renewcommand{\lowTEB}{{\tt lowTEB}}
\newcommand{\lowP}{{\tt lowP}}

\newcommand{\highell}{\ifmmode {\rm high}-\ell \else high-$\ell$\fi}
\newcommand{\hiell}{\ifmmode {\rm high}-\ell \else high-$\ell$\fi}
\newcommand{\lowell}{\ifmmode {\rm low}-\ell \else low-$\ell$\fi}

\newcommand{\vhl}{\ifmmode {\rm very-high}-\ell \else very-high-$\ell$\fi}
\renewcommand{\highL}{\vhl}
\renewcommand{\planckTT}{\planckTTonly\dataplus\lowP}

\newcommand{\hlp}{\texttt{Hillipop}}

\newcommand{\refeq}[1]{Eq.~\ref{eq:#1}}

\def\q#1{\lq{#1}\rq}

\newcommand{\cl}{\ensuremath{C_\ell}}
\newcommand{\clTT}{\ensuremath{C_\ell^{TT}}}

\newcommand{\clpp}{\ensuremath{C_\ell^{\Phi}}}

\newcommand{\lnAs}{\ln( 10^{10} \As)}
\newcommand{\bflike}{\lowTEB}
\newcommand{\bigO}[1]{\ensuremath{\mathcal{O}(#1)}}

\newcommand{\chimin}{\ensuremath{\chi^2_\textrm{min}}}

\newcommand{\AT}{\ensuremath{\mathcal{A}_T}}

\usepackage{ulem}
 
\begin{document}

\title{Relieving tensions related to the lensing of
the cosmic microwave background temperature power spectra} 
\titlerunning{Relieving tensions related to $\Alens$}

\author{F. Couchot, S. Henrot-Versill\'e, O. Perdereau,
  S. Plaszczynski\thanks{Corresponding author: \href{mailto:plaszczy@lal.in2p3.fr}{plaszczy@lal.in2p3.fr}}, B. Rouill\'e d'Orfeuil, M. Spinelli and M. Tristram}
\authorrunning{F. Couchot et al.}

\institute{Laboratoire de l'Acc\'el\'erateur Lin\'eaire, Univ. Paris-Sud, CNRS/IN2P3, Universit\'e Paris-Saclay, Orsay, France}

\abstract{
The angular power spectra of the cosmic microwave background (CMB) temperature anisotropies
reconstructed from \planck\ data seem to 
present \q{too much} gravitational lensing distortion. 
This is quantified by the control parameter $\Alens$ that should be
compatible with unity for a standard cosmology. 
With the \CLASS\ Boltzmann
solver and the profile-likelihood method, for this
parameter we measure  a 2.6$\sigma$ shift from 1 using the \planck\
public likelihoods.
We show that, owing to strong correlations with the reionization optical
depth $\tau$ and the primordial perturbation amplitude $\As$, a
$\sim2\sigma$ tension on $\tau$ also appears between the results obtained with the low ($\ell\leq 30$) 
and high ($30<\ell\lesssim 2500$) multipoles likelihoods. 
With \hlp, another \highell\ likelihood built from \planck\ data,
this difference is lowered to $1.3\sigma$.
In this case, the $\Alens$ value is still in disagreement with unity by
$2.2\sigma$, suggesting a non-trivial effect of the correlations between
cosmological and nuisance parameters.

To better constrain the nuisance foregrounds parameters, 
we include the \vhl\ measurements of the Atacama Cosmology Telescope 
(ACT) and South Pole Telescope (SPT) experiments and  obtain $\Alens = 1.03 \pm 0.08$.
The \hlp+ACT+SPT likelihood estimate of the optical depth is
$\tau=0.052\pm{0.035,}$ which is now fully compatible with the \lowell\
likelihood determination.
After showing the
robustness of our results with various combinations, we investigate
the reasons for this improvement that results from a better determination of the whole set of foregrounds parameters.
We finally provide
estimates of the \lcdm\ parameters with our combined CMB data likelihood.
}



\maketitle


\section{Introduction}
The $\Alens$ control parameter attempts to  measure the degree of
lensing of the cosmic microwave background (CMB)
power spectra. From a set of cosmological parameters ($\Omega$), a Boltzmann
solver, such as \CLASS\ \citep{class} or
\CAMB\ \citep{camb}, computes
the angular power spectra of the temperature/polarization
anisotropies $\cl(\Omega)$ and of the CMB  lensing potential $\clpp(\Omega)$. The latter is
then used to compute the distortion of the CMB spectra by the
gravitational lensing \citep{Blanchard1987}, which
redistributes the power across multipoles
while preserving the brightness in a non-trivial way
\citep[e.g][]{LewisChallinor}: $\{\cl(\Omega),\clpp(\Omega)\}\to \tilde{\cl}(\Omega)$. 
As originally proposed in \citet{Alens}, 
a phenomenological parameter, $\Alens$,  that re-scales the lensing potential, is introduced. 
This modifies the standard scheme into :  $\{\cl(\Omega),\Alens\cdot
\clpp(\Omega)\}\to \tilde{\cl}(\Omega,\Alens)$. 
Sampling the likelihood, with this parameter left free, gives access to
two interesting pieces of information:
\begin{enumerate}
\item from the $\Alens$ posterior distribution, one can check the consistency of the
  data with the model; it should be compatible with 1.0 for a standard cosmology.
\item by marginalizing over $\Alens$, one can study the impact of
  neglecting (to first-order) the lensing information contained in the CMB spectra.
\end{enumerate}

Since its first release, the \Planck\ Collaboration reports
a value of the $\Alens$ parameter that is discrepant with one by more than 2$\sigma$. The full-mission
result, based on both a high and \lowell\ likelihood \citep[][hereafter \params]{planck2014-a15}, is
\begin{equation}
\label{eq:planckAlens}
  \Alens = 1.22\pm 0.10 \quad \onesig{\planckTTonly+\lowP}
\end{equation}
(all quoted errors are 68\% CL intervals).
As shown later, a profile likelihood analysis, as the one in 
\pip, rather points to a $2.6\sigma$ discrepancy. 

This \q{tension} may indicate a problem either on the model or the data side.
The only solution for the model is to modify the computation of the geodesic deflection, i.e. to modify standard GR \citep{BHu:2015,Valentino:2015}.
For the data, since \planck\ maps undergo a complicated treatment
\citep[for an overview see][]{planck2014-a01}, one cannot exclude
small residual systematic effects that could  impact the details of
the likelihood function in a different way from one implementation
to the other. 


The anomalously high $\Alens$ value directly affects the measurement of two \lcdm\ parameters, the reionization optical depth $\tau$ and the primordial scalar perturbations amplitude $\As$. 
Indeed, in the \highell\ regime, only the $\AT\equiv\As e^{-2\tau}$
combination is constrained by the temperature power spectra
amplitude. However this degeneracy is broken by the lensing distortion of the CMB anisotropies since $\clpp\propto\As$ (more perturbations induce more lensing) so that both $\As$ and $\tau$ finally get  constrained.

The aim of this work is twofold. First, to clarify the connection 
between the $\Alens$ tension with unity and the one that
also appears on $\tau$ between the \planck\ public high and \lowell\ likelihoods, 
and also to show that this effect may be related to the details of the
nuisance parametrization in the likelihoods.
By using \hlp, a \highell\ likelihood that is built from \planck\ data, and
better constraining the astrophysical foregrounds {and the high-$\ell$
  part of the CMB spectrum} with the high-angular resolution data from
ACT and SPT, we show that one can obtain a more self-consistent picture of the \lcdm\ parameters. 
Section~\ref{sec:planck} provides an in-depth discussion about
$\Alens$ using the \planck\ baseline likelihoods, \plik\ and \bflike,
and makes  the link with the determination of $\tau$ explicit.
Section~\ref{sec:hlp} then recalls the main differences between \plik\
and \hlp\ and discusses the first results with the latter.
Then Sect.~\ref{sec:act_spt} describes how the inclusion of the ACT and SPT data
was performed and, after various checks, discusses how their inclusion
impacts the \hlp\ results. Finally Sect.~\ref{sec:lcdm} discusses the results on the \lcdm\
parameters using \hlp\ in combination with other likelihoods.


\section{The \Planck\ $\Alens$ tension (and related parameters)}
\label{sec:planck}
\subsection{\Planck\ likelihoods}
\label{sec:terminology}
\Planck's baseline uses two different likelihood codes addressing different multipole ranges \citep[see][hereafter \liket]{planck2014-a13}:
\begin{enumerate}
\item the \highell\ likelihood 
    (\plik) is a Gaussian likelihood that acts in the multipole range
    $\ell\in[30,2500]$. Data consist of a 
  collection of angular power spectra that are derived from
  cross-correlated \planck\
  $100,143,$ and $217$~GHz high frequency maps. 
  For the results in this paper, we only use the temperature likelihood.
\item the \lowell\ likelihood (\bflike) is a pixel-based likelihood that essentially relyies
   on the \planck\ low frequency instrument $70$~GHz maps for polarization and on a component-separated map using all \planck\ frequencies for temperature. It acts in the $\ell \in [2,29]$ range.
\end{enumerate}
In the \liket\ terminology, \q{$\planckTTonly$} refers to the combination of both the high and the \lowell\
temperature likelihoods. In this case, only the TT component of \bflike\ is used.
The \q{$\planckTTonly\dataplus\lowP$} notation combines \plik\ to the full \bflike\ likelihood, which we   label explicitly in this paper as \plik+\bflike.

In the following, we will make use of the publicly available \plik\
likelihood code (\verb=plik_dx11dr2_HM_v18_TT.clik=)\footnote{available in the Planck Legacy Archive (PLA): \\\PLA} with
the Gaussian priors on nuisance parameters suggested by the \planck\ collaboration\footnote{\href{http://wiki.cosmos.esa.int/planckpla2015}{http://wiki.cosmos.esa.int/planckpla2015}}, where the information on foregrounds from the ACT and SPT data is
propagated by a single SZ prior (\citetalias[][Sect. 2.3.1]{planck2014-a15}).

\subsection{Boltzmann solver}
\label{sec:boltzmann}
The results derived in this paper make use of the Boltzmann equations solver
\CLASS\footnote{\wwwclass} 
while \Planck's published results were derived using \CAMB\footnote{\wwwcamb}.
Both softwares have been compared previously \citep{classIII} and have produced spectra in excellent agreement when using their respective  high precision
settings. More recentl, it was noticed  in \params\ that sampling from 
any of them gives very compatible results on \lcdm\ cosmological parameters.
The precise estimate of the $\Alens$ parameter is more challenging
since one is dealing with sub-percent effects on the spectra and thia requires
extra care about differences between both softwares.

For this purpose, we sampled the \plik+\bflike\ likelihoods in the
\lcdm+$\Alens$ model using \CLASS\ \texttt{v2.3.2} and obtain results
almost identical to the published ones. In particular we measure
\begin{equation}
  \Alens=1.24\pm{0.10} \quad \onesig{\plik+\lowTEB,~\CLASS/MCMC}.
\label{eq:plik_Alens_class_MC}
\end{equation}
The tiny difference with respect to \refeq{planckAlens} can be traced down to a \bigO{1$\muk$} difference on the \highell\ part of the
TT spectra (see Appendix \ref{app:classvscamb}), but we consider that
this general 
agreement is sufficient to perform reliable estimations. All further
results will be derived consistently using \CLASS.

\subsection{Profile likelihoods}
\label{sec:profile}
In this paper, we  also often make use of a statistical methodology based on
profile-likelihoods for reasons that will be clearer in Sect. \ref{sec:plik_tau}.
For a given parameter $\theta$, we perform several multi-dimensional minimizations of the 
$\chi^2\equiv-2\ln{\mathscr{L}}$ function. Each time $\theta$
is fixed to a given $\theta(i)$ value, a minimization is performed
with respect to all the other parameters, and the $\chimin(i)$ value is
kept. The curve interpolated through the
$\{\theta(i),\chi^2_\text{min}(i)\}$ points and offset to 0, is known as the $\theta$ profile-likelihood:
$\Delta\chi^2(\theta)$. We note that from the very construction procedure, the
solution at the minimum of the profile always coincides with the
complete best-fit solution, i.e the maximum likelihood estimate
(MLE) of all the parameters. A genuine
 68\% CL interval is obtained by thresholding the profile 
at one even in non-Gaussian cases \citeg{jamesbook}. 

This statistical method, an alternative to Monte-Carlo Markov
Chain (MCMC) sampling, was discussed in \pip\ and also used in \liket.
Building a smooth profile from \planck\ data is computationally challenging since this approach requires an extreme precision on the $\chimin$ solution, typically better than 0.1 for values around $10^4$. 
This goal can be achieved using the \texttt{Minuit} software\footnote{\wwwminuit} together with 
an increase of the \CLASS\ precision parameters.
For the analysis presented in this paper, we have further refined the
procedure described in \pip, as explained in Appendix~\ref{app:minuit}. 

This procedure leads to a so-called confidence interval
\citeg{jamesbook}. In the frequentist approach, this represents a
statement on the data: when repeating the experiment many times, the
probability for the reconstructed interval to cover the true value is
68\%. The Bayesian approach, as implemented through a MCMC method,
leads to what is generally referred to as a \q{credible interval}
derived from the probability density function of the true value. 
In most cases (in particular Gaussian) both intervals are very similar.
However, in some cases (typically so-called banana-shaped 2D posteriors), these intervals may differ significantly \citep{Porter96}.
In this case, the mode (or mean) of the posterior distribution does not necessarily match the best fit solution and the existence of a difference between both values indicates what is referred to as likelihood volume effects.
In this paper, we will mainly focus on the details of the region around the maximum likelihood and x will thus use  the profile-likelihood method consistently.

\subsection{$\Alens$ revisited}
\label{sec:alens}

\begin{figure}[!ht]
\centering
\includegraphics[width=88mm]{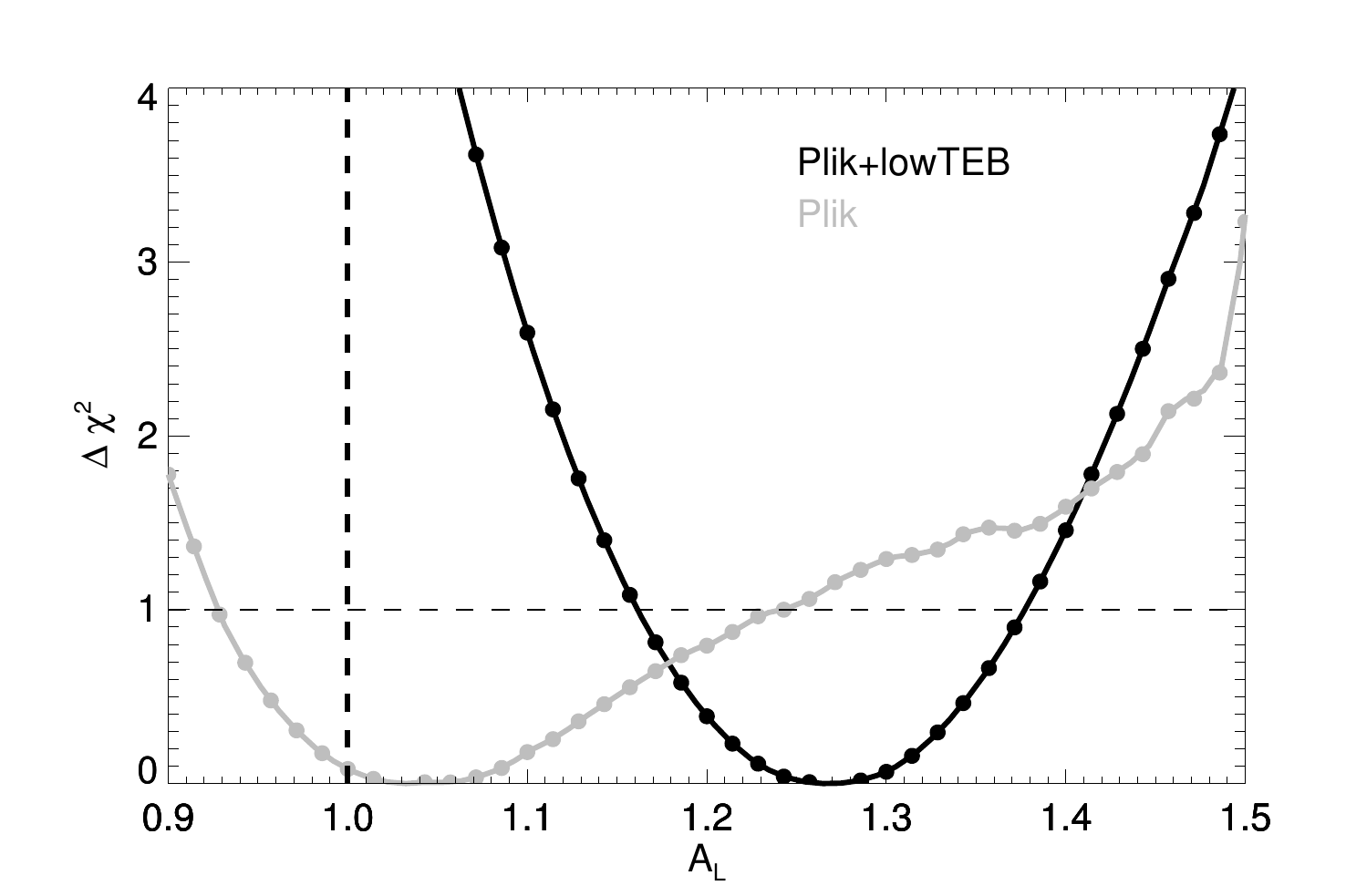}
\caption{Profile-likelihoods of the $\Alens$ parameter reconstructed
  from the \plik\ \hiell\ likelihood alone (in grey) and when
  adding the \bflike\ one (in black).
  The vertical dashed line recalls the expected \lcdm\ value.}
\label{fig:Plik_Alens}
\end{figure}

We first build the profile-likelihood for $\Alens$
(Fig.~\ref{fig:Plik_Alens}) and measure
\begin{equation}
\label{eq:Plik_Alens_class}
  \Alens=1.26_{-0.10}^{+0.11}\quad \onesig{\plik+\lowTEB,~\CLASS/profile}.
\end{equation}

The shift with respect to \refeq{plik_Alens_class_MC} quantifies the size of the volume effects in the MCMC projection.
It is of the same order of magnitude as the $\CAMB \to \CLASS$ transition seen in Sect.~\ref{sec:boltzmann}.
Using high-precision settings, we therefore find $\Alens$ at $2.6\sigma$ from 1.0.

Figure~\ref{fig:Plik_Alens} also shows that the \plik-alone likelihood (in grey) gives
\begin{equation}
  \Alens=1.04_{-0.10}^{+0.20}\quad \onesig{\plik,~\CLASS/profile},
\end{equation}
which is compatible with 1.0. This difference from the \planck\ baseline result (\refeq{Plik_Alens_class}) seems to come from a tension between the low and \hiell\ likelihoods.
Moreover, using a prior of the kind $\tau=0.07\pm0.02$ (as in \liket) leads to $\Alens = 1.16 \pm 0.09,$ which goes in the same direction as \plik+\lowTEB.
This connection with $\tau$ will be discussed in the following section.

\subsection{High- vs. \lowell\ likelihood results on $\tau$ and $A_s$}
\label{sec:plik_tau}

We further investigate the high vs. low-$\ell$ likelihood tensions from
the point of view of two other parameters that are strongly correlated
to $\Alens$: the reionization optical depth $\tau$ and the scalar
perturbation amplitude $\As$.

\begin{figure}[!ht]
\centering
\includegraphics[width=88mm]{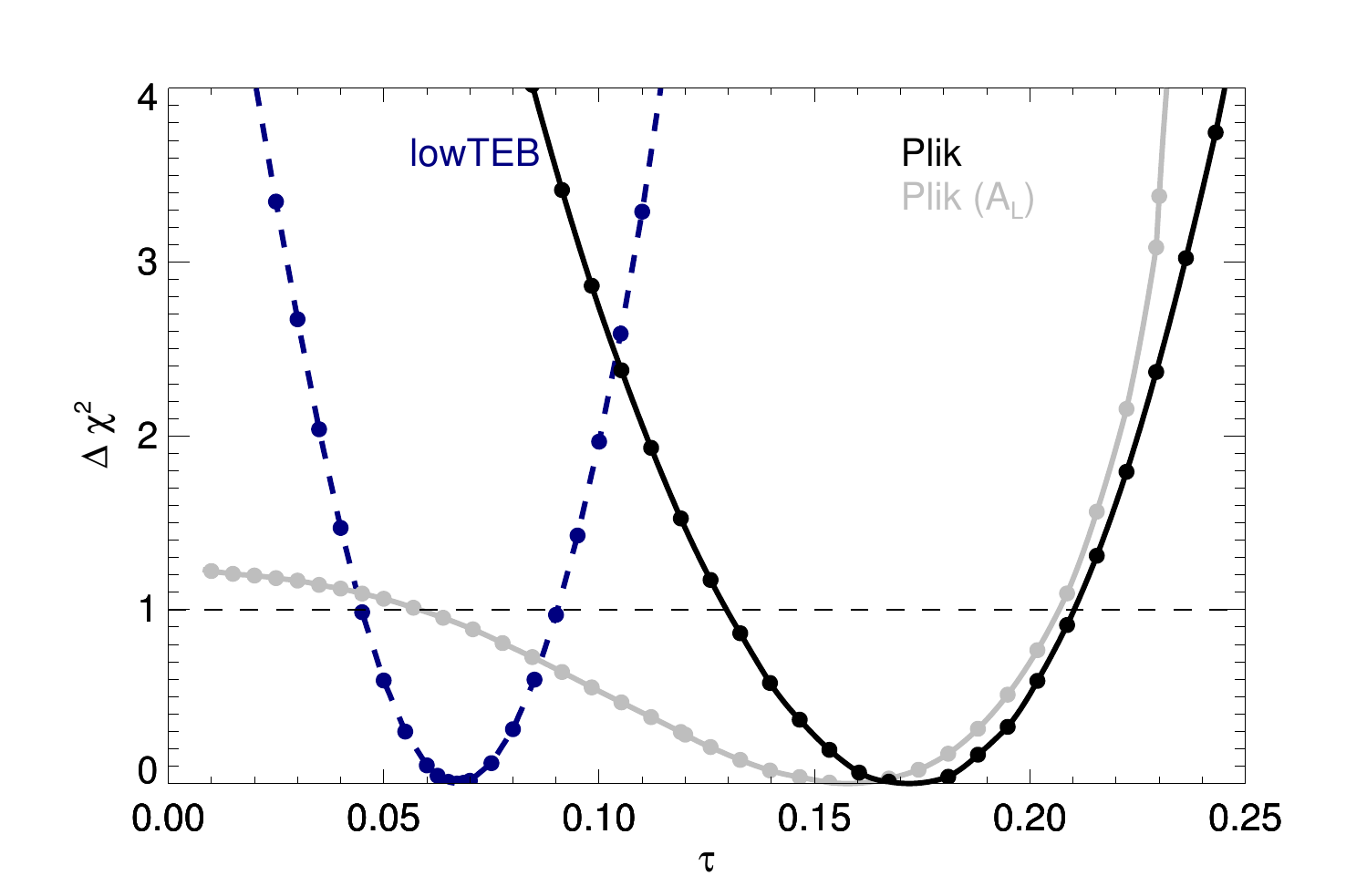}
\caption{High vs. \lowell\ \planck\ likelihood constraints on $\tau$. 
  The \hiell\ result is obtained with \plik\ (only) and is shown in black
  while the \lowell\ one is in dashed blue.
  Both are obtained within the \lcdm\ model. The grey profile shows the
  result for \plik\ when $\Alens$ is left free in the fits.}
\label{fig:plik_tauprof}
\end{figure}

Figure~\ref{fig:plik_tauprof} shows, in black, the $\tau$ profile-likelihood reconstructed with \plik\ only, which gives
\begin{equation}
  \label{eq:plik_tau}
  \tau=0.172^{+0.038}_{-0.042} \quad \onesig{\plik}.
\end{equation}

This is higher than the maximum of the posterior reported in \liket\
(Fig.~45) that is around 0.14 and is partly due to a volume effect 
(Sect. \ref{sec:profile}) and partly because of the \CLASS/\CAMB\
difference highlighted in Appendix~\ref{app:classvscamb}.
Using consistently \CLASS\ and the same methodology, our result is
$2.2\sigma$ away from the $\tau$ determination 
with the \bflike\ likelihood for which the profile-likelihood gives
the same results as the MCMC marginalization (\liket):
\begin{align}
\label{eq:lowtau}
  \tau&=0.067^{+0.023}_{-0.021} \quad \onesig{\lowTEB},
\end{align}
represented as a blue line on Fig. \ref{fig:plik_tauprof}.

Without fixing $\Alens$ to 1 (grey curve on
Fig. \ref{fig:plik_tauprof}) the constraint on $\tau$ is much weaker, which
illustrates the fact that the lensing of the CMB anisotropies in the
\hiell\ likelihood is the main contributor to the $\tau$ measurement. 
Some constraining power still remains in particular for large $\tau$
values: this is due to the fact that the degeneracy between $\As$ and $\tau$ is broken for large $\tau$
when the reionisation bump at low $\ell$ enters the multipole range of \plik\ ($\ell > 30$)~\citep[see][]{huwhite97}.

The discrepancy highlighted in Fig.~\ref{fig:plik_tauprof} is directly related to the $\Alens$ problem
(Fig. \ref{fig:Plik_Alens}), but is simpler to study. 
The \hiell\ likelihood requires a large $\tau$ value that is in tension with the \lowell-based result. 
In the $\Alens$ test (Fig. \ref{fig:Plik_Alens}) one combines both
likelihoods.\lowTEB\ pulls $\tau$ down. To match the spectra amplitude ($\AT$), the \hiell\
likelihood pulls $\As$ down. Then $\Alens$, being fully
anti-correlated to $\As$ (since $\clpp\propto\Alens\As$), 
shifts to adjust the lensed model to the  data again.

Because of the \AT\ degeneracy, the \plik-only estimate
of $\As$ is also expected to be high. Indeed from a similar
profile-likelihood analysis we obtain
\begin{equation}
  \lnAs=3.270_{-0.078}^{+0.058} \quad \onesig{\plik},
\end{equation}
again discrepant by more than 2$\sigma$ with the results from \plik+\bflike, $3.089\pm0.036$ (\params).

In summary, the \plik\ \hiell\ likelihood alone converges to a consistent solution, $\Alens\simeq 1$, but with
large $\tau$ and $\As$ values. Constraining $\tau$ down by adding 
the \bflike\ likelihood (or a low prior) is compensated in the fits by increasing $\Alens$ to match
the data.
To investigate the stability of those results, we will now use another
\Planck\ \hiell\ likelihoods.


\section{The \hlp\ likelihood}
\label{sec:hlp}
\subsection{Description}

\begin{figure*}[!ht]
\centering
\includegraphics[width=\textwidth,bb=50 40 670 540,clip]{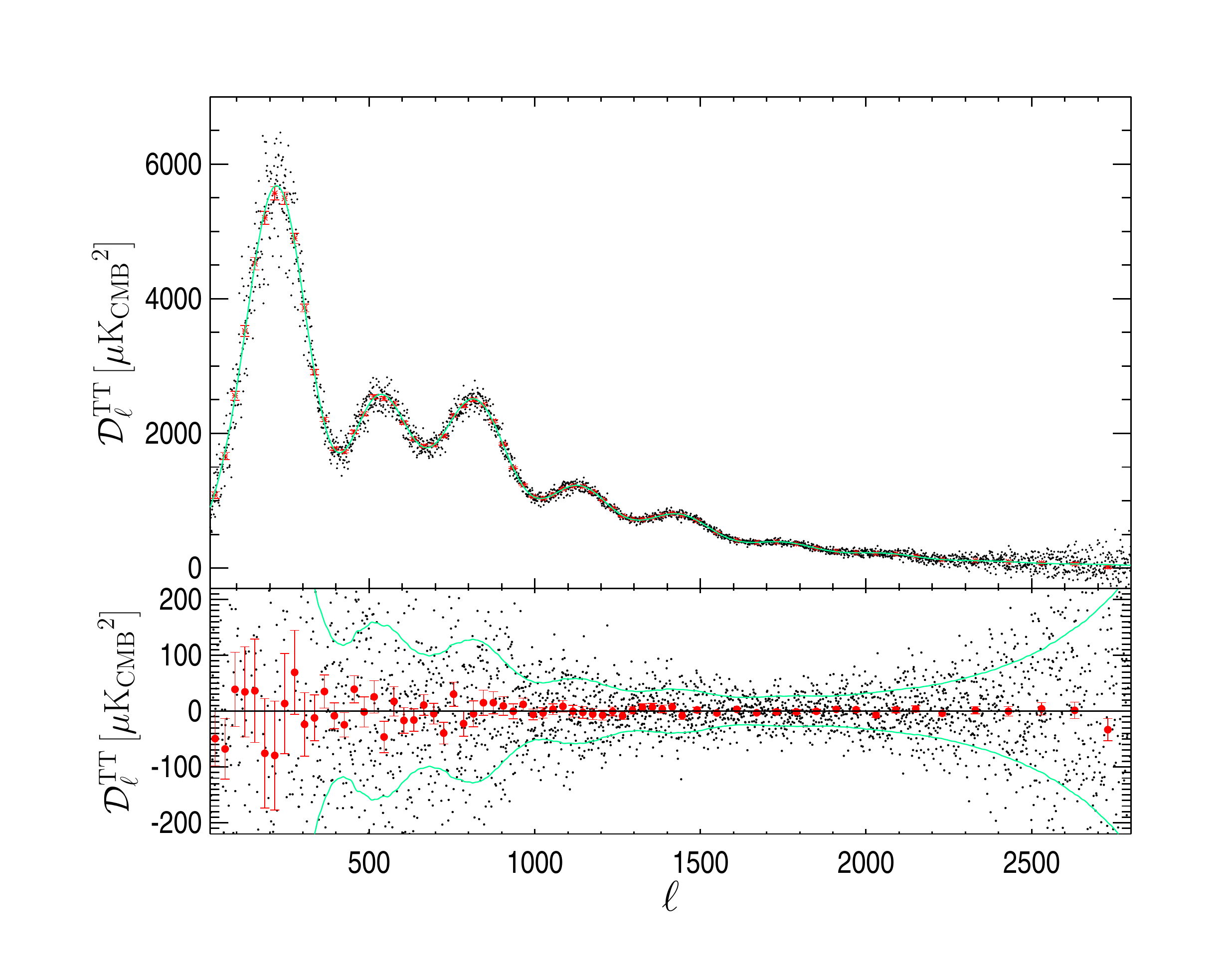}
\caption{\hlp\ foreground-subtracted combined power-spectrum ($\mathcal{D}_\ell=\ell(\ell+1)C_\ell/2\pi$) at each multipole
  (black points) and binned (red points) with respect to the best-fit
  model. The bottom plot shows the residuals. The green line shows the standard deviation as estimated from
  the covariance matrix.}
\label{fig:TTspec}
\end{figure*}

\hlp\ is one of the \planck\ \hiell\ likelihoods developed for the
2015 data release and is shorty described in \liket. Similarly to \plik, it is a Gaussian likelihood based on
cross-spectra from the HFI 100, 143, and 217 GHz maps. 
The estimate of cross-spectra on data is performed using
\texttt{Xpol}, a generalization of the \texttt{Xspec} algorithm \citep{Xspect} to polarization.
Figure~\ref{fig:TTspec} shows the combined TT spectrum with respect to the best-fit model that will be deepened later on.

The differences with \plik\ were mentioned in \liket. The most significant  are:
\begin{itemize}
\item we use all the 15 half-mission cross-spectra built
  from the 100, 143, and 217~GHz maps while \plik\ uses only five of them;
\item we apply inter-calibration coefficients at the map level,
  resulting in five free parameters (one is fixed) while \plik\ uses
  two at the spectrum level;
\item we use point-sources masks that were obtained from a refined
  procedure that extracts Galactic compact structures;
\item as a result, our
  galactic dust component follows closely and is parametrized by the power law discussed in
  \citet{planck2014-XXX}; 
\item we use foreground templates derived from
  \citet{planck2013-pip56} for the CIB, and \citet{planck2014-a28} for
  the SZ;
\item we use all multipole values (i.e., do not bin the spectra).
\end{itemize}

Using \hlp\  leads to \lcdm\ estimates that are very  compatible with the
other \planck\ ones but on $\As$ and $\tau$ (\liket, Sect.~4.2).
Using a prior on $\tau$ of $0.07\pm0.02$, we obtain with \hlp\ $\tau=0.075\pm0.019$, while
\plik\ gives a higher value $\tau=0.085\pm0.018$ (\liket).
Given the relation between $\tau$ and $\Alens$ discussed in Sect.~\ref{sec:plik_tau}, we
can therefore expect different results on $\Alens$.

\begin{table}[!ht]
\begin{center}
\begin{tabular}{llc}
\hline
\hline
Name & Definition & Prior (if any) \\
\hline
\multicolumn{3}{c}{Instrumental}\\
\hline
$c_0$ & map calibration (100-hm1) &  $1.000 \pm 0.002$ \\  
$c_1$ & map calibration (100-hm2) &  $1.000 \pm 0.002$ \\  
$c_2$ & map calibration (143-hm1) &  fixed to 1. \\  
$c_3$ & map calibration (143-hm2) &  $1.0000 \pm 0.002$ \\  
$c_4$ & map calibration (217-hm1) &  $1.0025 \pm 0.002$ \\  
$c_5$ & map calibration (217-hm2) &  $1.0025 \pm 0.002$ \\  
$A$ & absolute calibration & $1.0000 \pm 0.0025$ \\
\hline
\multicolumn{3}{c}{Foreground modelling} \\
\hline
$A_{\rm PS}^{100 \times 100}$   & PS amplitude in TT  (100x100 GHz) &\\
$A_{\rm PS}^{100 \times 143}$   & PS amplitude in TT (100x143 GHz) &\\
$A_{\rm PS}^{100 \times 217}$   & PS amplitude in TT (100x217 GHz) &\\
$A_{\rm PS}^{143 \times 143}$   & PS amplitude in TT (143x143 GHz) &\\
$A_{\rm PS}^{143 \times 217}$   & PS amplitude in TT (143x217 GHz) &\\
$A_{\rm PS}^{217 \times 217}$   & PS amplitude in TT (217x217 GHz) &\\

$\mathbf{A_{SZ}}$ & scaling for the tSZ template &\\
$\mathbf{A_{CIB}}$ & scaling for the CIB template &$ 1.00 \pm 0.20 $ \\
$\mathbf{A_{kSZ}}$ & scaling for the kSZ template &\\
$\mathbf{A_{SZxCIB}}$ & scaling parameter for the & \\ 
 & cross-correlation between kSZ and CIB &\\
$A_{\rm dust}^{\rm TT}$ & scaling parameter for the dust in TT & $1.00 \pm 0.20$\\
\hline
\hline
\end{tabular}
\caption{Nuisance parameters for the \hlp\ likelihood and Gaussian prior used during
the likelihood maximization. The calibration factors are taken from \citet{planck2014-a09}
and the amplitude of the dust template extrapolated from the 353\GHz\
\citep{planck2013-pip56}. A prior on $A_{\rm CIB}$ is applied to
relieve some degree of degeneracy among foregrounds when ACT+SPT data
are not used. We have put in boldface the parameters that are
  common to the ACT and SPT likelihoods in the combined fit (see Sect.~\ref{sec:act_spt}).}
\label{tab:hlp_nuisance}
\end{center}
\end{table}


\subsection{Results}
\label{sec:hlp_results}

\begin{figure}[!ht]
\centering
\includegraphics[width=88mm]{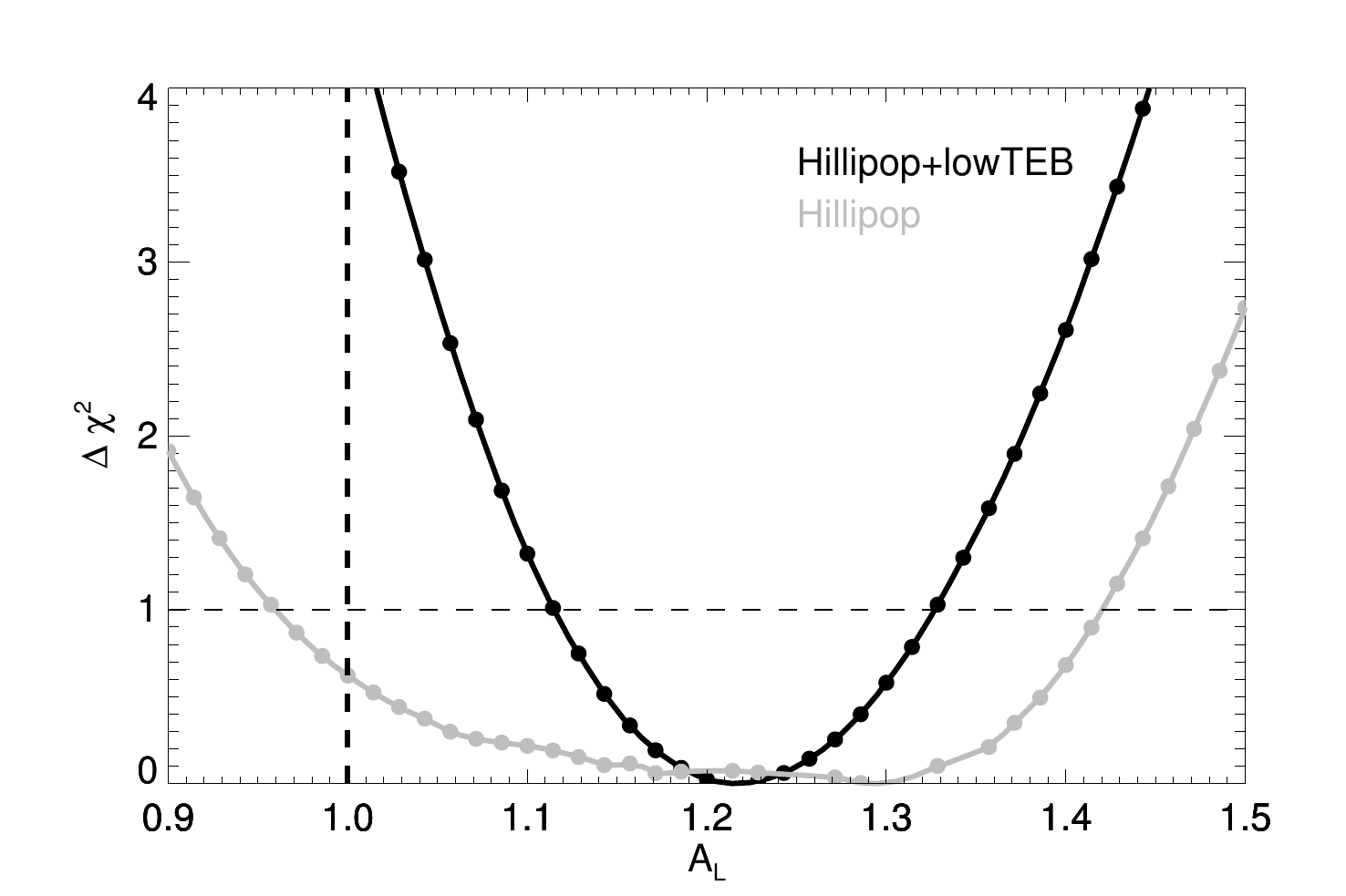}
\caption{Profile-likelihoods of the $\Alens$ parameter reconstructed from the \hlp\ likelihood alone (in grey) and when adding \bflike\  (in black). The vertical dashed line recalls the expected $\Lambda$CDM value.}
\label{fig:hlp_Alens}
\end{figure}

The profile-likelihoods of $\Alens$ derived from  \hlp\ with and without \bflike\ is shown in Fig.~\ref{fig:hlp_Alens}. 
The \hlp-alone profile is minimum near $\Alens=1.30$ but is very broad : a 68\% CL interval goes from  .96 up to 1.42. 
We therefore conclude that \hlp\ alone does not give a strict constraint on $\Alens$.  
In combination with \bflike,\, using the same procedure as described in Sect.~\ref{sec:alens}, we obtain
\begin{align}
\label{eq:hlp_Alens}
\Alens=1.22_{-0.10}^{+0.11} \quad \onesig{\hlp+\lowTEB}.
\end{align}

This is slightly lower than the result obtained with \plik\ (\refeq{Plik_Alens_class})
but still discrepant with one by about $2\sigma$.

\begin{figure}[!htb]
\centering
\includegraphics[width=88mm]{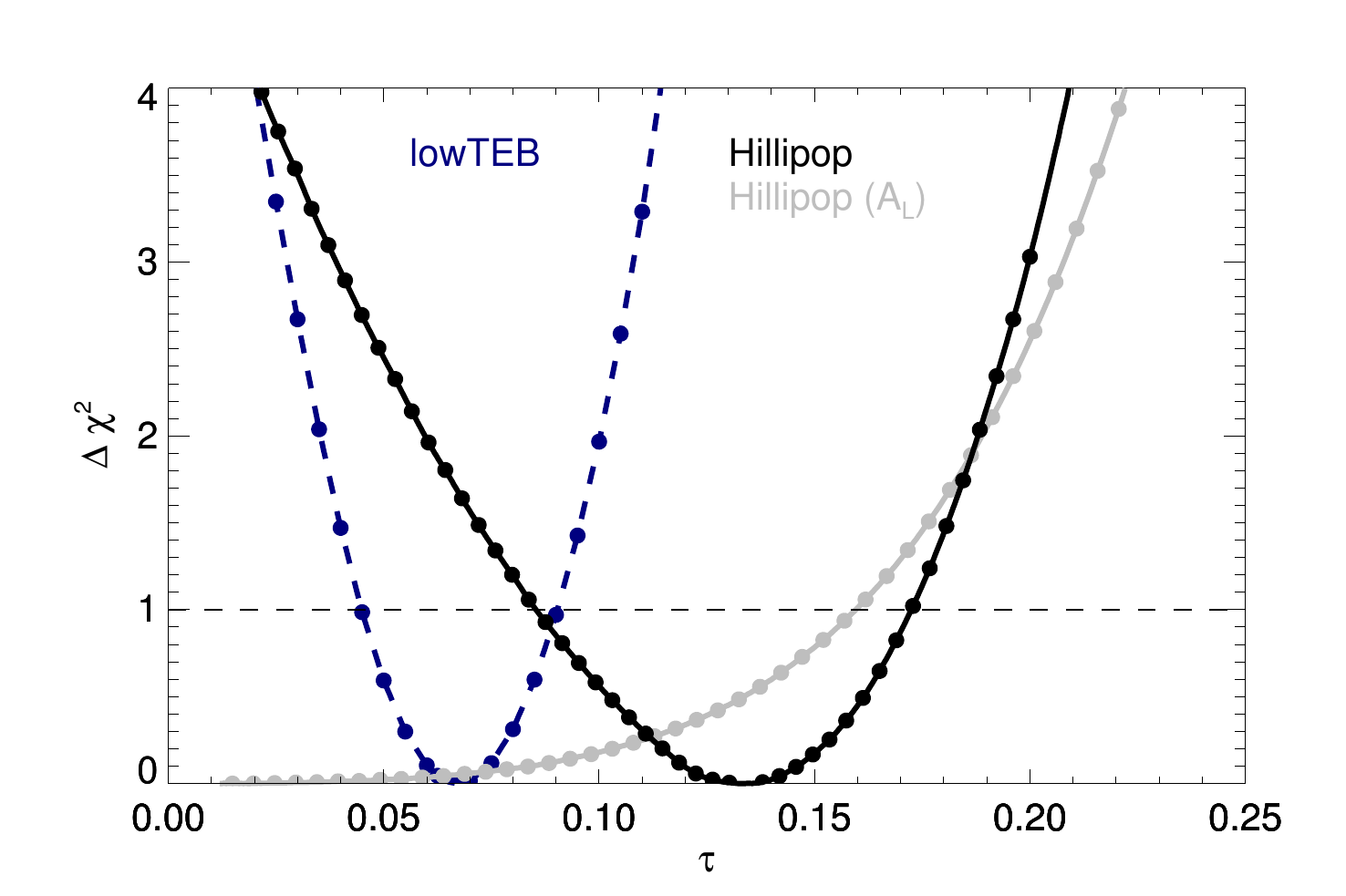}
\caption{Profile-likelihoods of the $\tau$ parameter using only the
  \hiell\ \hlp\ likelihood: \lcdm+$\Alens$ free in the fits (in grey) and \lcdm\ with fixed $\Alens=1$ (in
black).}
\label{fig:hlp_tauprof}
\end{figure}

Within the \lcdm\ model, Fig.~\ref{fig:hlp_tauprof} compares \hlp\ vs. \bflike\ results on $\tau$. The \hlp\ profile on $\tau$ gives
\begin{align}
\label{eq:tauhlp}
  \tau&=0.134^{+0.038}_{-0.048} \quad \onesig{\hlp}.
\end{align}
This is lower than the \plik\ result with similar error bars (\refeq{plik_tau}) and lies within 1.3$\sigma$ of the \lowell\ measurement (\refeq{lowtau}).
In the \lcdm+$\Alens$ case, \hlp\ only gives an upper limit. The difference with \plik\ $\tau$ profile (Fig.~\ref{fig:plik_tauprof}) is the sign of different correlations between $\Alens$ and $\tau$ in these likelihoods.



One of the difference between the two likelihoods is in the definition of the foreground models. 
Moreover, with \planck\ data only, the accuracy on the foreground parameters is weak (especially for SZ and CIB amplitudes).
In the next section, we will use the \vhl\ datasets. This  both adds constraints on lensing through the high multipoles and  better determines the foregrounds parameters and possibly modifies the non-trivial correlations between nuisance and cosmological parameters.


\section{Adding \highL\ data to constrain the foregrounds}
\label{sec:act_spt}

\subsection{ Datasets} 
\paragraph{ Atacama Cosmology Telescope.}

We use the final ACT temperature power spectra presented in
\cite{Das:2014}. These are 148$\times$148, 148$\times$218, and 218$\times$218
power spectra built from observations performed on two different sky areas
(south and equatorial) and during several seasons, for
multipoles between 1000 and 10000 (for 148$\times$148), and 1500 to 10000
otherwise.

\paragraph{South Pole  Telescope.}

We use two distinct datasets from SPT. 

The higher $\ell$ part, dubbed {\tt SPT\_high}, uses results, described in 
 \cite{Reichardt:2012}, from observations
at 95, 150, and 220\GHz\ from the SPT-SZ survey. Their cross-spectra cover the $\ell$ range
between 2000 and 10000. These measurements were calibrated using WMAP~7yr data. 
A more recent analysis from the complete $\sim$2500\,$\mathrm{deg}^2$ area of the  SPT-SZ
survey is presented in \cite{George:2014}, dubbed {\tt SPT\_high2014} hereafter.  
In this later release,  cross spectra cover a somewhat broader  $\ell$ range, 
between 2000 and 13000. Both sets of cross spectra are however quite similar, but the later 
comes with a covariance matrix that includes calibration uncertainties. Its use makes 
our work harder since it was calibrated on the \planck\ 2013 data, which in turn had 
a calibration offset of~1\% (at the map level) with respect to the
\planck\ 2015 spectra. We thus prefer to use 
 \cite{Reichardt:2012} dataset as a baseline in our analyses, with free calibration parameters to match other datasets.
We have checked that all results presented in this paper are stable when switching to \cite{George:2014}, in which case we have to 
set strict priors on recalibration parameters owing to the form of the associated covariance matrix.

We also include the \citet{Story:2012} dataset, dubbed {\tt SPT\_low},
 consisting of a 150\GHz\ power spectrum which ranges from $\ell=650$  to 3000.
 Some concerns were raised in \citet{planck2013-p11} about the compatibility of this
 dataset with \planck\ data. The tension was
 actually traced to be with the \WMAP+SPT cosmology and the \planck\ and {\tt SPT\_low} power spectra
 were found to be broadly consistent with each other \citep{planck2013-p11}.
 As will be shown later, we do not see any sign of tension between the
 \planck\ 2015 data and the {\tt SPT\_low} dataset, nor any reason to
 exclude it.

\subsection{Foregrounds modelling}

For the \vhl\ astrophysical foregrounds,
we chose to use a model as coherent as possible with what has been set-up for \hlp,
i.e., the same templates for tSZ, kSZ, CIB, and tSZ$\times$CIB.
Since they have been computed for the 
\planck\ frequencies and bandpasses, we have to extrapolate them to the ACT and SPT respective effective 
frequencies and bandpasses. For tSZ, we scale the template with the
usual $f_{\nu}=x\coth{x/2}-4$ function (where $x=h\nu/k_{\mathrm{B}}T_{\mathrm{CMB}}$), using the effective frequencies for the SZ spectral distribution  
given in  \cite{Dunkley:2013}. 
For CIB and tSZ$\times$CIB, we start from templates in $Jy^2.sr^{-1}$ in the
IRAS convention ($\nu I(\nu)=\mathrm{cste}$ spectrum) for \planck\ effective
frequencies and bandpasses. For CIB, we use the conversion factors
from  \planck\ to the ACT/SPT  effective frequencies and bandpasses, assuming the
\cite{Bethermin:2012} SED for the CIB combined with unit conversion factors to $K_{CMB}$, for the ACT and
SPT bandpasses \citep{GLAgache:2014}. These factors are given in Table~\ref{tab:vhl_conv_factors}.

 \begin{table}[h!]     
 \begingroup
 \openup 5pt
 \newdimen\tblskip \tblskip=5pt
 \nointerlineskip
 \vskip -3mm
 \footnotesize
 \setbox\tablebox=\vbox{
     \newdimen\digitwidth
     \setbox0=\hbox{\rm 0}
     \digitwidth=\wd0
     \catcode`*=\active
     \def*{\kern\digitwidth}
     \newdimen\signwidth
     \setbox0=\hbox{+}
     \signwidth=\wd0
     \catcode`!=\active
     \def!{\kern\signwidth}
 \halign{
 \hbox to 0.5in{#}\tabskip=10pt& \hfil$#$\hfil &\hfil $#$ \hfil &\hfil $#$ \hfil&\hfil $#$ \hfil \cr
 \noalign{\doubleline}
 \omit \hfil Dataset \hfil &\omit\hfil Channel \hfil &\omit\hfil
 $MJy.sr^{-1}/K_{CMB}$ \hfil &\omit\hfil HFI freq.\hfil & \omit
 Conversion\cr
 & \omit \hfil (GHz) \hfil & & \omit \hfil (GHz) \hfil \cr
 \noalign{\vskip 3pt\hrule\vskip 5pt}
ACT & 148 & 401.936 & 143 & 0.85 \cr
    & 218 & 485.311 & 217 & 1.056 \cr 
 \noalign{\vskip 3pt\hrule\vskip 5pt}
SPT & 95 & 234.042 & 100 & 1.090 \cr
 & 150 & 413.540 & 143 & 0.7688 \cr
 & 220 & 477.017 & 217 & 1.061 \cr
 \noalign{\vskip 5pt\hrule\vskip 3pt}
 } 
 } 
 \endPlancktable
 \endgroup
\caption{\label{tab:vhl_conv_factors} Conversion factors used for the foreground template extrapolation to ACT and SPT bandpasses with the CIB SED.}
\end{table}

For the  tSZ$\times$CIB component of the ($\nu_1\times\nu_2$) cross-spectrum (from the ACT or SPT dataset), we scale the nearest HFI cross-spectrum ($\nu_1^P\times\nu_2^P$)  using the ratio 
\begin{equation}
S_{\nu_1,\nu_2}=\frac{f_{\nu_1}C_{\nu_2}+f_{\nu_2}C_{\nu_1}}{f_{\nu^P_1}C_{\nu^P_2}+f_{\nu^P_2}C_{\nu^P_1}}
\label{eq:snu1nu2}
,\end{equation}
and then convert it to $K_{CMB}$ using the factors computed from the for the ACT and
SPT bandpasses, as above. 
This scaling applies 
at the 15\% level for the HFI cross-frequency templates, choosing the 143$\times$143 one as a reference. 

In addition, 
a few more specific templates have been added to each datasets: 
\begin{itemize}
\item Point sources : to mask resolved point sources, ACT and SPT used their own settings to match each instrument's sensitivity and angular resolution. The unresolved point source populations in each case are thus different, so we introduce extra nuisance parameters to model them.
We model the unresolved point source components in the ACT and SPT spectra with one amplitude $A^{\nu_1 \times \nu_2}_{PS}$ parameter per cross-spectrum. Consequently, this  introduces six nuisance parameters for the ACT, six for the {\tt SPT\_high,} and one for the {\tt SPT\_low} datasets, respectively (see Table~\ref{tab:vhl_nuisance}).

\item Galactic dust : following \cite{Dunkley:2013} and \cite{Das:2014}, we model the dust contribution in the ACT power spectra as a power law 
\begin{equation}
\mathcal{D}_{\ell}^{dust}(i,j)\ =\ A_{\rm dust}^{\rm ACT}\left(\frac{\ell}{3000}\right)^{-0.7}\left(\frac{\nu_i\nu_j}{\nu_0^2}\right)^{3.8}\left[\frac{g(\nu_i)g(\nu_j)}{g(\nu_0)^2}\right]
.\end{equation}
We therefore introduce  two nuisance parameters, one for each part of the ACT dataset, and set the reference frequency $\nu_0$ at 150\GHz.

For the SPT datasets, following \cite{Reichardt:2012}, we use a fixed template, with amplitudes 0.16, 0.21, and 2.19 $\mu K_{CMB}^2$ at 95, 150, and 218\GHz, respectively and an $\ell^{-1.2}$ spatial dependency. 
\end{itemize}

\subsection{Likelihoods}

\begin{figure*}[!ht]
\centering
\includegraphics[width=\textwidth,viewport=50 40 670 540,clip]{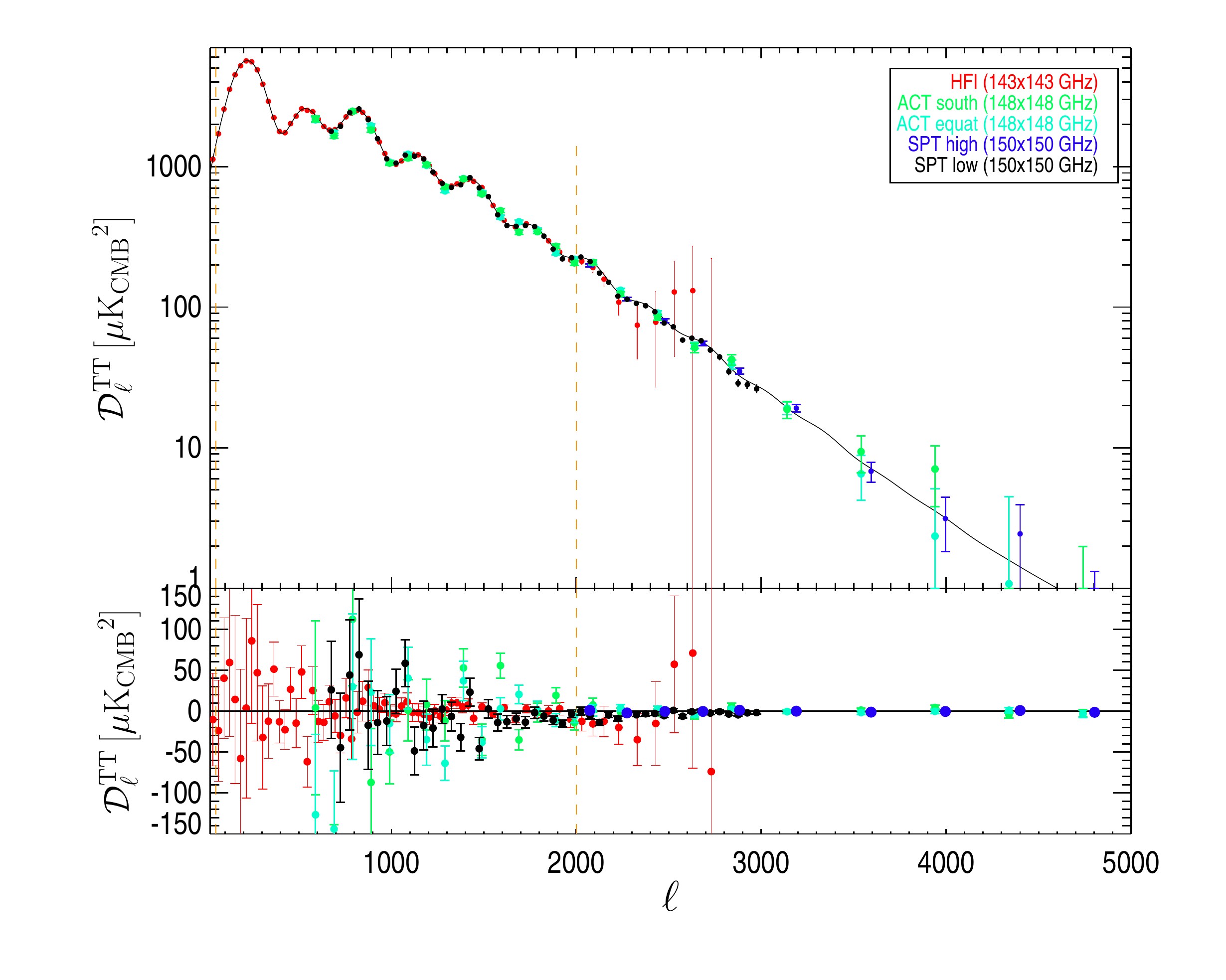}
\caption{Foreground-subtracted CMB cross-spectra of all the experiments
  used at $\simeq 150$ GHz. The solid line is the best fit of the
  \hlp+VHL combination and is subtracted to obtain the bottom
  residual plot. The \hlp\ likelihood uses individual multipoles up to 2000 and 
  window functions have been accounted for in the \vhl\ data.}
\label{fig:TTSpec_w_VHL_143x143_xlin}
\end{figure*}

We compute one likelihood for each of the five \vhl\ datasets following
the  method described in \citet{Dunkley:2013}. We use the respective
published window functions to bin the (CMB + foregrounds)  model, and
the released covariance matrices to compute the likelihood. In all cases, these include beam uncertainties. Since we combine different datasets, we introduced nine additional nuisance parameters to account for their relative calibration uncertainties (at map level). Figure~\ref{fig:TTSpec_w_VHL_143x143_xlin} shows a comparison of all the
 foreground-subtracted CMB spectra for the 150$\times$150 component (which is almost common to all experiments), 
 and Table \ref{tab:vhl_nuisance} summarizes the characteristics of the datasets we use
 in the VHL likelihoods.

 A detailed inspection of all cross-spectra per frequency and
 component has been performed and does not reveal any
 inconsistency with the \planck\ data. More details are given in
 Appendix~\ref{app:fg}.

 \begin{table}[h!]     
 \begingroup
 \openup 5pt
 \newdimen\tblskip \tblskip=5pt
 \nointerlineskip
 \vskip -3mm
 \footnotesize
 \setbox\tablebox=\vbox{
     \newdimen\digitwidth
     \setbox0=\hbox{\rm 0}
     \digitwidth=\wd0
     \catcode`*=\active
     \def*{\kern\digitwidth}
     \newdimen\signwidth
     \setbox0=\hbox{+}
     \signwidth=\wd0
     \catcode`!=\active
     \def!{\kern\signwidth}
 \halign{
 \hbox to 1.2in{$#$\leaderfil}\tabskip=5pt& \hfil$#$\hfil &\hfil $#$ \hfil &\hfil $#$ \hfil \cr
 \noalign{\doubleline}
 \omit \hfil Dataset \hfil &\omit\hfil Freq (GHz) \hfil &\omit\hfil
 \#spectra \hfil &\omit\hfil \#nuisances \hfil \cr
 \noalign{\vskip 3pt\hrule\vskip 5pt}
{\tt SPT\_low} & 150  & 1 & 2\cr
{\tt SPT\_high} & 95, 150, 220 & 6 & 9 \cr
{\tt ACT~ south/equat} & 148, 218 & 6 & 12 \cr
 \noalign{\vskip 5pt\hrule\vskip 3pt}
 } 
 } 
 \endPlancktable
 \endgroup
 \caption{%
Summary of the characteristics of the \vhl\ data used in this analysis. Each experiment's likelihood includes map calibrations
 and residual point source levels, which result in a number of additional nuisance parameters shown in the last column.
 In combined fits, the SZ and CIB foreground templates are common with \hlp.
}
\label{tab:vhl_nuisance} 
\end{table}

\subsection{First results and global consistency check\label{ssect:firstvhlres}}

\begin{figure}[!ht]
\centering
\includegraphics[width=88mm]{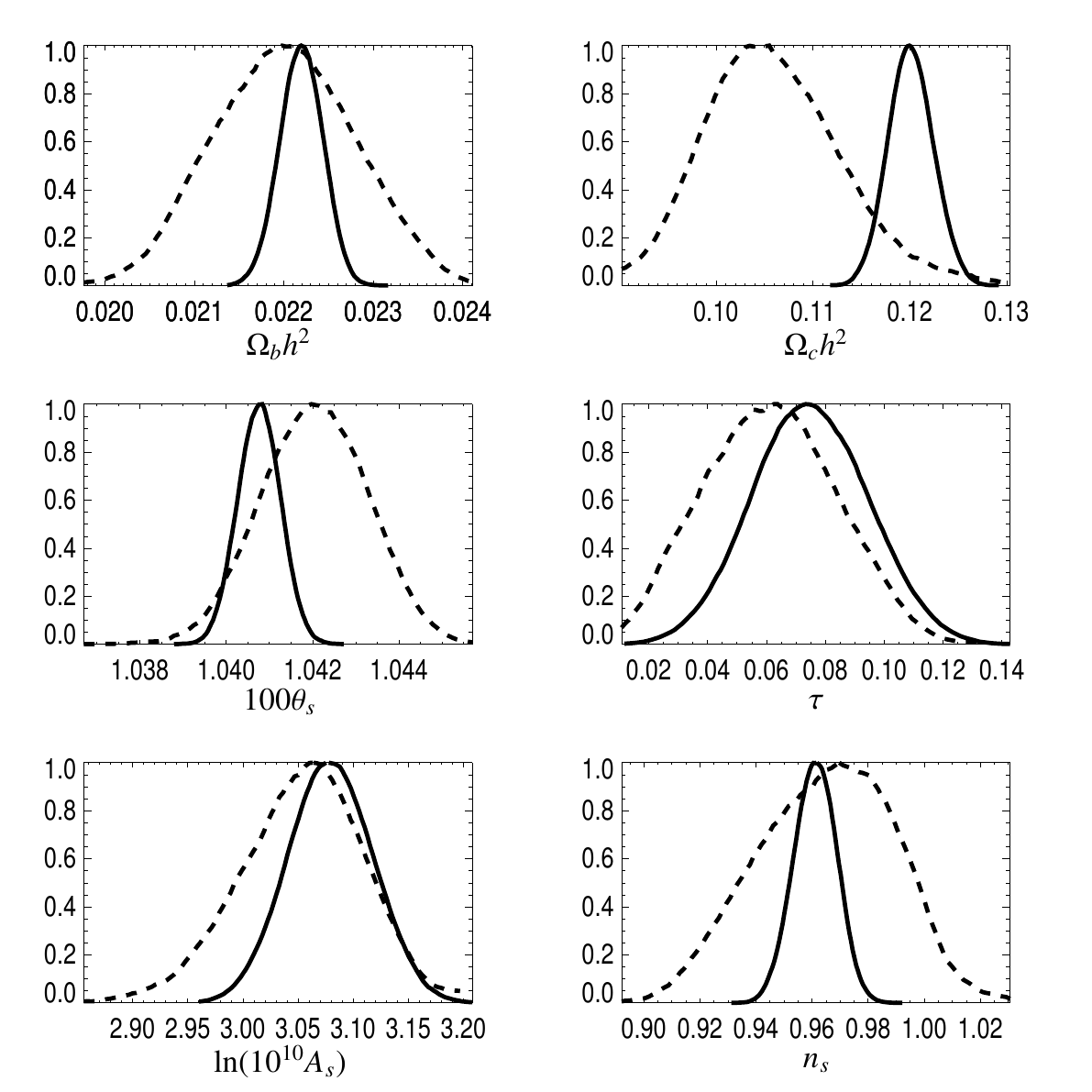}
\caption{Posterior distributions of
  the cosmological parameters obtained by sampling the \hlp\ (soled line) and
  ACT+SPT (dashed) likelihoods independently.  
  A prior of $\tau=0.07\pm0.02$ is used in both
  cases. For clarity, we only show the
  cosmological parameters, but all nuisances are sampled.}
\label{fig:vhl_alone}
\end{figure}

We first check that the combination of the ACT+SPT
likelihoods (hereafter VHL) gives results consistent with
\hlp. We therefore sample  the \hlp\ and
VHL likelihoods independently and compare their \lcdm\ estimates
in Fig.~\ref{fig:vhl_alone}. 
Both datasets lead to similar cosmological parameters.
However, an accurate measurement of $\Omega_bh^2$ and $\Omega_ch^2$ requires a precise determination of the relative amplitudes of the CMB acoustic peaks for which the \vhl\ datasets are less sensitive.
This is reflected by the width of the posteriors shown in Fig.~\ref{fig:vhl_alone}.

The $\chi^2$  contributions to the \hlp+VHL+\lowTEB\ best fit
of each of the \vhl\  datasets 
are 58 /47(d.o.f.),  77 /90, and 651/710 for the
{\tt SPT\_low}, {\tt SPT\_high,} and {\tt ACT} datasets, respectively. 
None of these individual values 
indicate strong tension between the likelihood parts. We note that in all cases, the covariance matrices provided by the ACT and SPT groups, and used to compute the $\chi^2$, 
include non negligible, non-diagonal elements. For example, the visual impression from the {\tt SPT+low} residuals, shown on  Fig.~\ref{fig:fgdetails_SPTlow} is not excellent, but the 
$\chi^2$ from this part of the fit given above has a PTE of 13\%, which
is perfectly acceptable.

\subsection{$\Alens$ and $\tau$ results}
\begin{figure}[!ht]
\centering
\includegraphics[width=88mm]{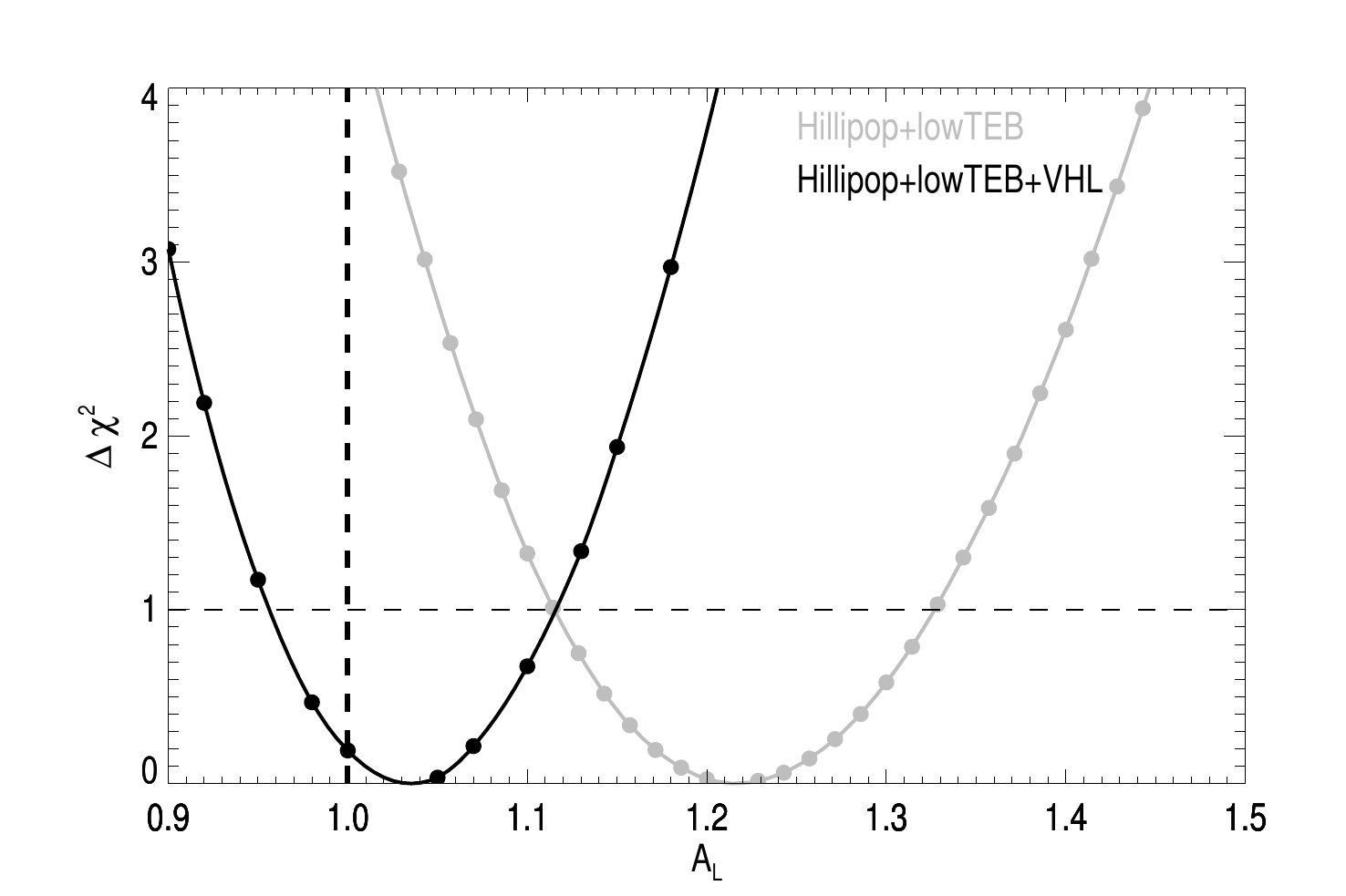}
\caption{Profile-likelihood for $\Alens$ reconstructed
  from the \hlp+\lowTEB\ likelihood (in grey) and adding the \vhl\
  ACT and SPT data (VHL) discussed in the text (in black).}
\label{fig:hlp_vhl_Alens}
\end{figure}

As a second step, we perform the same analysis as in Sects.~\ref{sec:planck} and~\ref{sec:hlp_results},
adding the VHL likelihood to \hlp\ and  consider the $\Alens$
profile-likelihood (with \lowTEB) in Fig.~\ref{fig:hlp_vhl_Alens}.
The result becomes
\begin{equation}
  \label{eq:Al_vhl}
   \Alens=1.03\pm0.08\quad \onesig{\hlp+\lowTEB+VHL},
\end{equation}
now fully compatible with one.

\begin{figure}[!ht]
\centering
\includegraphics[width=88mm]{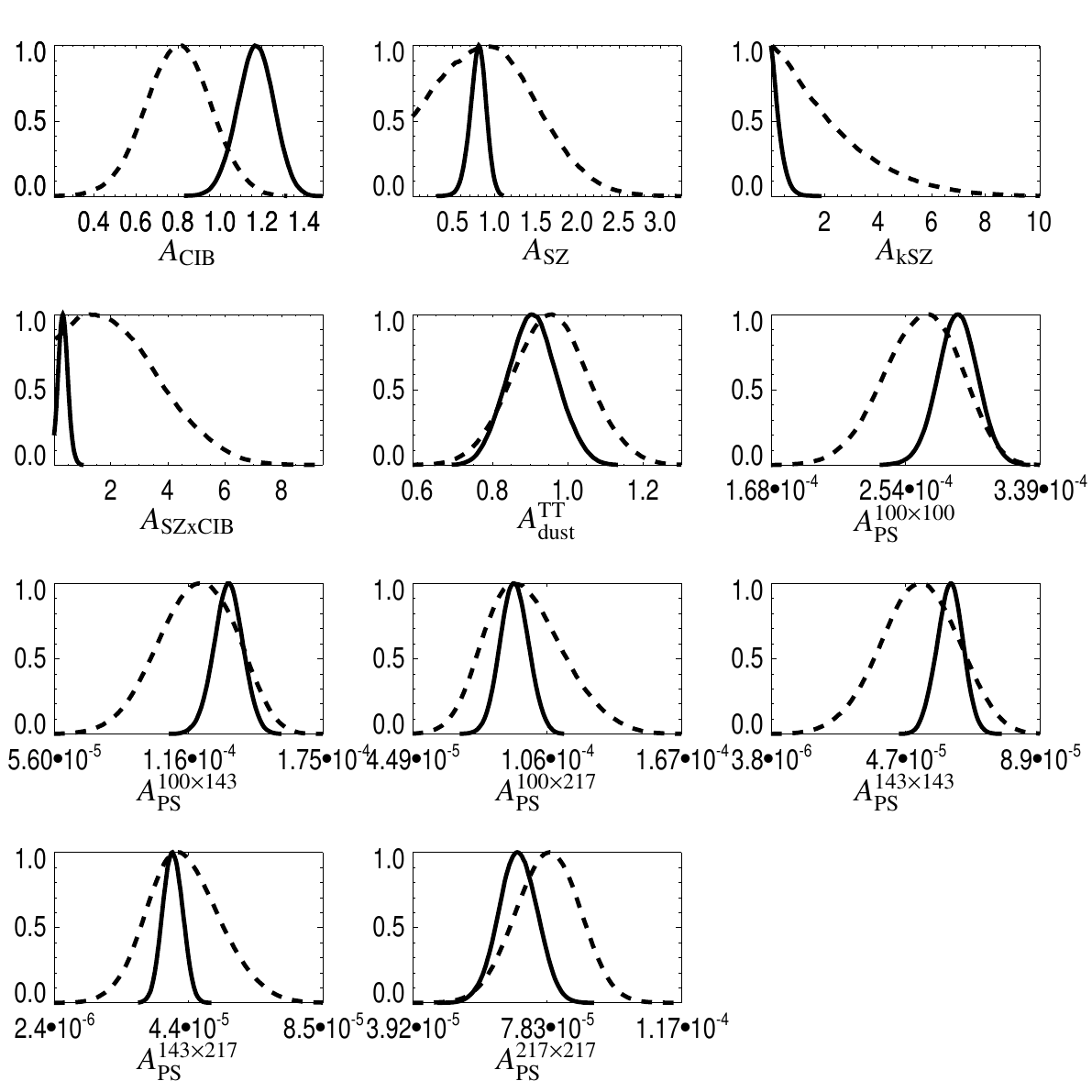}
\caption{Posterior distributions of the foreground \hlp\ parameters
  with (solid line) and without (dashed line) the VHL likelihood. The
  definition of each parameter can be found in Table~\ref{tab:hlp_nuisance}.}
\label{fig:fgcomp}
\end{figure}

As seen in Sect. \ref{sec:lcdm}, the inclusion of the VHL data does not greatly change  the cosmological parameters but, as expected, strongly constrains all the foregrounds including, through correlations, the ones specific to \planck\ data (dust and point source amplitudes). This is shown in Fig.~\ref{fig:fgcomp}.

We note that all \hlp\ nuisance amplitudes (but the point sources) represent coefficients scaling foreground templates: it is remarkable that, after adding the VHL likelihood, they all lie reasonably (at least those for which we have the sensitivity) around one, which is a strong support in favor of the coherence of the foregrounds description.

\begin{figure}[!ht]
\centering
\includegraphics[width=88mm]{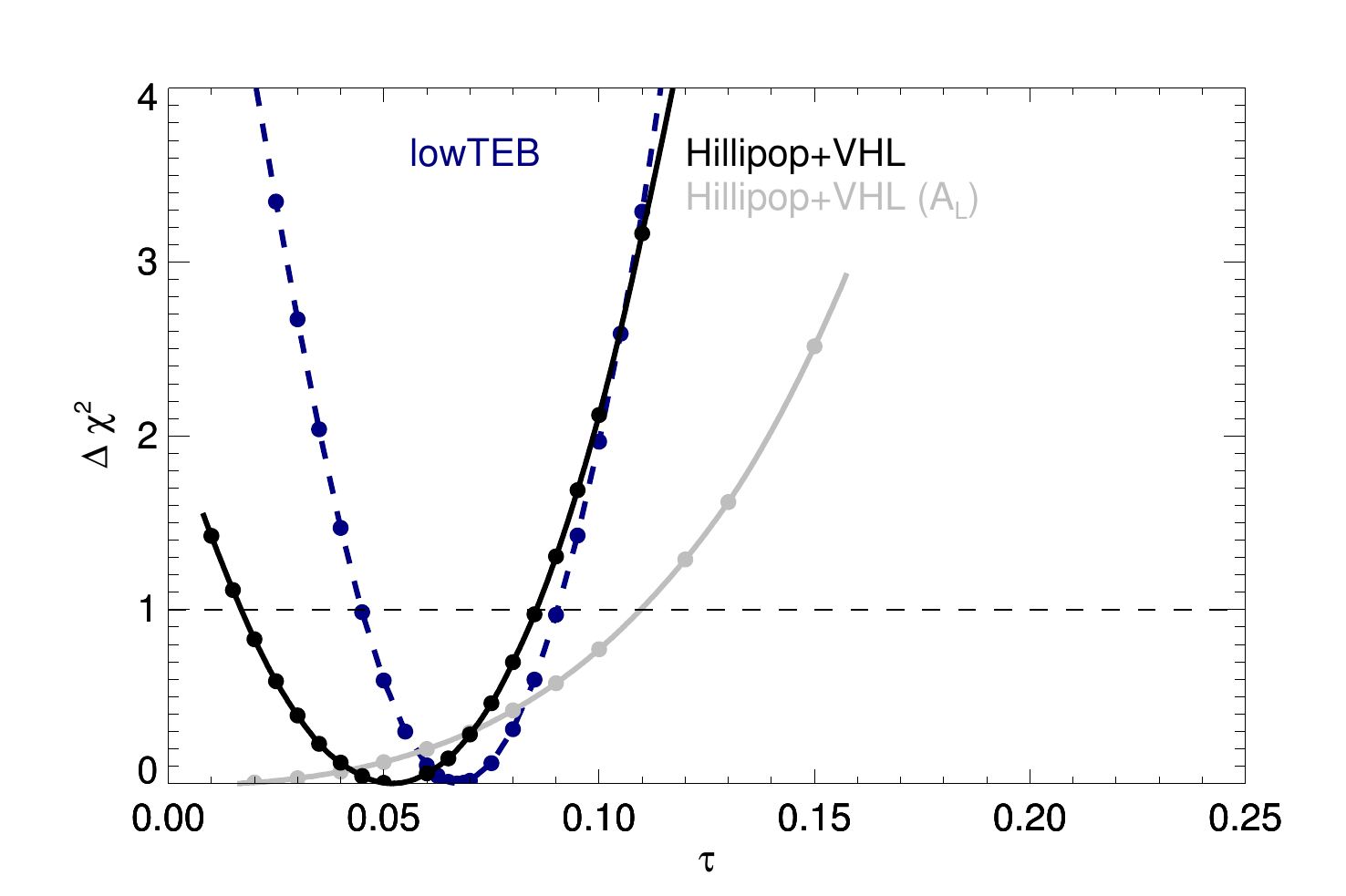}
\caption{\hlp+VHL and \bflike\ likelihood constraints on $\tau$.
 The grey profile shows the result for \hlp+VHL when $\Alens$ is left free in the fits.
  The \bflike\ one is in dashed blue.
}
\label{fig:hlp_vhl_tauprof}
\end{figure}

For $\tau$, combining the VHL with the \hlp\ likelihood removes any sign of tension with \bflike\ as shown in Fig. \ref{fig:hlp_vhl_tauprof}, and we obtain
\begin{align}
\label{eq:tauhlpvhl}
  \tau&=0.052\pm{0.035} \quad \onesig{\hlp+VHL},
\end{align}
which is in excellent agreement with the \lowTEB\  measurement (\refeq{lowtau}).

\subsection{Robustness of the results}
\label{sec:syst}

We tested a large number of configurations of the VHL dataset to
establish whether the improvement comes from a particular one. Results
are presented in Table~\ref{tab:Alens_vhls}.

 \begin{table}[h!]     
 \begingroup
 \openup 5pt
 \newdimen\tblskip \tblskip=5pt
 \nointerlineskip
 \vskip -3mm
 \footnotesize
 \setbox\tablebox=\vbox{
     \newdimen\digitwidth
     \setbox0=\hbox{\rm 0}
     \digitwidth=\wd0
     \catcode`*=\active
     \def*{\kern\digitwidth}
     \newdimen\signwidth
     \setbox0=\hbox{+}
     \signwidth=\wd0
     \catcode`!=\active
     \def!{\kern\signwidth}
 \halign{
 \hbox to 1.8in{$#$\leaderfil}\tabskip=10pt& \hfil$#$\hfil \cr
 \noalign{\doubleline}
 \omit \hfil Dataset \hfil&\omit\hfil $\Alens$ \hfil \cr
 \noalign{\vskip 3pt\hrule\vskip 5pt}
 {\tt None} & 1.22\pm0.11\cr
 {\tt SPT\_low} & 1.16\pm0.10\cr
 {\tt SPT\_high} & 1.12 \pm0.10\cr
 {\tt ACT} & 1.19\pm0.10\cr
 {\tt SPT\_low}+{\tt SPT\_high} & 1.02\pm0.08\cr
 {\tt SPT\_low}+{\tt ACT} & 1.09\pm0.09 \cr
 {\tt SPT\_high}+{\tt ACT} & 1.12\pm0.09 \cr
 {\tt SPT\_low}+{\tt SPT\_high}+{\tt ACT} & 1.03\pm0.08\cr
 \noalign{\vskip 5pt\hrule\vskip 3pt}
 } 
 } 
 \endPlancktable
 \endgroup
 \caption{%
Results on $\Alens$ using \hlp+\lowTEB\ and various dataset combinations on the
VHL side. For their exact definition see Sect. \ref{sec:act_spt}.}                    
\label{tab:Alens_vhls} 
\end{table}

It is difficult to draw  firm conclusions from this exercise, since the number of
extra nuisance parameters  varyies in each case (see Table. \ref{tab:vhl_nuisance}). 
The improvement on $\Alens$ is most significant when combining several
datasets, but also satisfactory as soon as one combines at least two of them. 
We note that all these results are highly correlated with each other, sinc they all make use of \hlp.
Even though central values very close to 1.0 may be preferable, it should be pointed out that
there are several combinations that are compatible with 1.0 at the 
$\sim1\sigma$ level. This may indicate that the better the constraint on the high end of the power spectrum, the better the constraint on the $\Alens$ control parameter 
(the lensing effect  on $C_{\ell}^{TT}$ is not only a smearing of
peaks and troughs but also a redistribution of power towards the high
$\ell$, above $\sim 3000$).

Since there is some overlap between {\tt SPT\_low} and {\tt SPT\_high} datasets, we expect some correlations between the {\tt SPT\_low} power-spectrum and, in particular, the 150\GHz\ spectrum of {\tt SPT\_high}. To check the impact on the results, we either removed the entire 150\GHz\ from {\tt SPT\_high} or the overlapping bins in multipole for each of the two datasets and re-ran the analysis.
In all cases the results were similar with the combined one. 
For example, when removing the bins at $\ell < 3000$ from the 150x150\GHz\ of {\tt SPT\_high}, we find $\Alens= 0.99\pm0.08$, which is in good agreement with unity and with the results reported in Table~\ref{tab:Alens_vhls}. More checks are described in Appendix~\ref{app:fg}.

Finally, we noticed that the {\tt SPT\_high} 220$\times$220 spectrum lies
slightly high with respect to the best fit (see Appendix~\ref{app:fg}).
Indeed, we get a better agreement in this case by increasing the contribution of the SPT dust amplitude.
To check the impact on $\Alens$, we re-run the analysis, multiplying the dust level for all SPT spectra, $A_{\rm dust}^{\rm SPT}$, 
by a factor of 3 and obtain
\begin{align}
\Alens&=1.13\pm 0.10 ~
\onesig{\hlp+\lowTEB+SPT\_high, $A_{\rm dust}^{\rm SPT}\times$3} \nonumber \\
\Alens&=1.04\pm0.08 ~
\onesig{\hlp+\lowTEB+VHL, $A_{\rm dust}^{\rm SPT}\times$3} \nonumber.
\end{align}
Compared with Table~\ref{tab:Alens_vhls},
the details of the SPT dust amplitude do not affect the final results.

\subsection{Where does the change on $\Alens$ come from?}
\label{sec:change}

Adding the ACT and SPT data lowered the $\Alens$ estimate. In this
section, we try  to pinpoint where the change came from.

As discussed in Sect. \ref{sec:terminology}, \planck\ included the
\vhl\ information in the \plik\ likelihood through a linear constraint
between the thermal and kinetic components of the SZ foreground.
When combining \hlp\ with the VHL likelihood, a similar correlation is observed
that in our units reads:
\begin{equation}
\label{eq:sz}
  A^\textrm{kSZ}+3.5A^\textrm{tSZ}=3.16\pm0.25
.\end{equation}

To check whether this correlation is sufficient to capture the essentials of the
\vhl\ information, we re-run the profile analysis adding to \hlp+\bflike\ only
the prior in \refeq{sz} and measure:
\begin{align}
\Alens=1.26_{-0.10}^{+0.12} \quad \onesig{\hlp+\lowTEB+SZ-cor}.
\end{align}

A comparison with \refeq{hlp_Alens} shows that, at least in our case, using this correlation
does not capture the complexity of the full covariance matrix.

\begin{figure}
  \centering
\includegraphics[width=88mm]{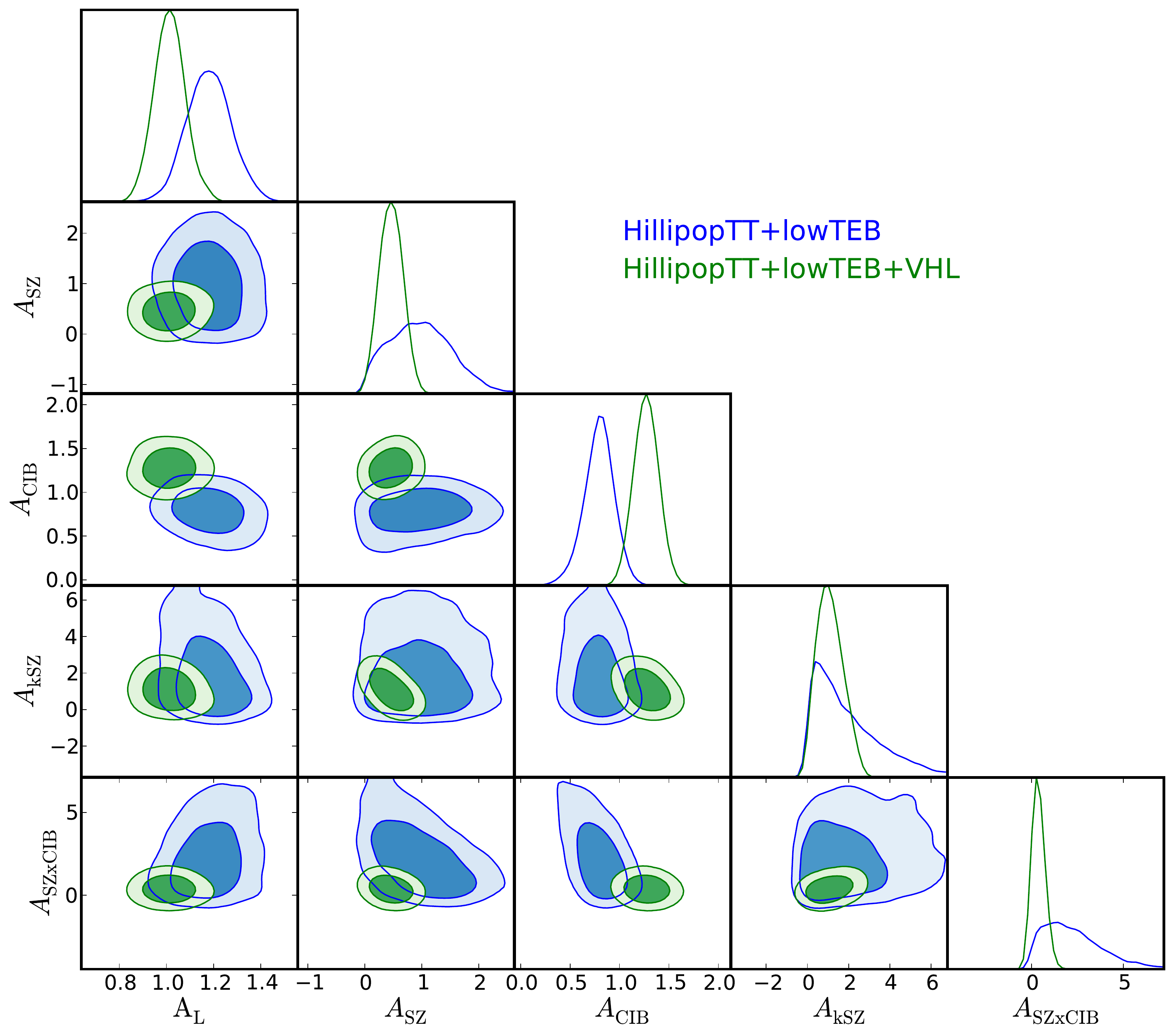}
\caption{\label{fig:MCMCAlens} Posterior distritions (68 and 95\%
  levels) obtained from sampling the \hlp+\lowTEB\ and
  \hlp+\lowTEB+VHL likelihoods. We 
  display the parameters most constrained by VHL.}
\end{figure}

So, to check if the change came from a better constraint 
over all the foregrounds, we perform the following measurement: we use the
\hlp+\lowTEB\ likelihood to determine $\Alens$ as in
Sect. \ref{sec:hlp_results} but fixing all the nuisance parameters to
the best-fit value of \hlp+\lowTEB+VHL likelihood. A profile-likelihood analysis
gives
\begin{align}
\Alens=1.09\pm0.08\quad \onesig{\hlp+\lowTEB, \text{fixed nuisances}}, \nonumber
\end{align}
compatible with unity at $\sim1\sigma$ as when using the \vhl\
data. We conclude that the shift for $\Alens$ seems to come from the
better determination of the foregrounds parameters.
To determine which particular foreground parameters impacts
the $\Alens$ shift, we run an MCMC analysis sampling all the
parameters (including $\Alens$) with the \hlp+\lowTEB\ and \hlp+\lowTEB +VHL
likelihoods.
The posterior distributions for $\Alens$ and the foregrounds parameters in common between \hlp\ and VHL are shown on Fig.~\ref{fig:MCMCAlens}. 
It is difficult to single out a particular
correlation and we conclude that the improvement comes from the overall better constraint of the foregrounds.


\section{Results on \lcdm\ parameters}
\label{sec:lcdm}
\begin{figure}[!ht]
\centering
\includegraphics[width=88mm]{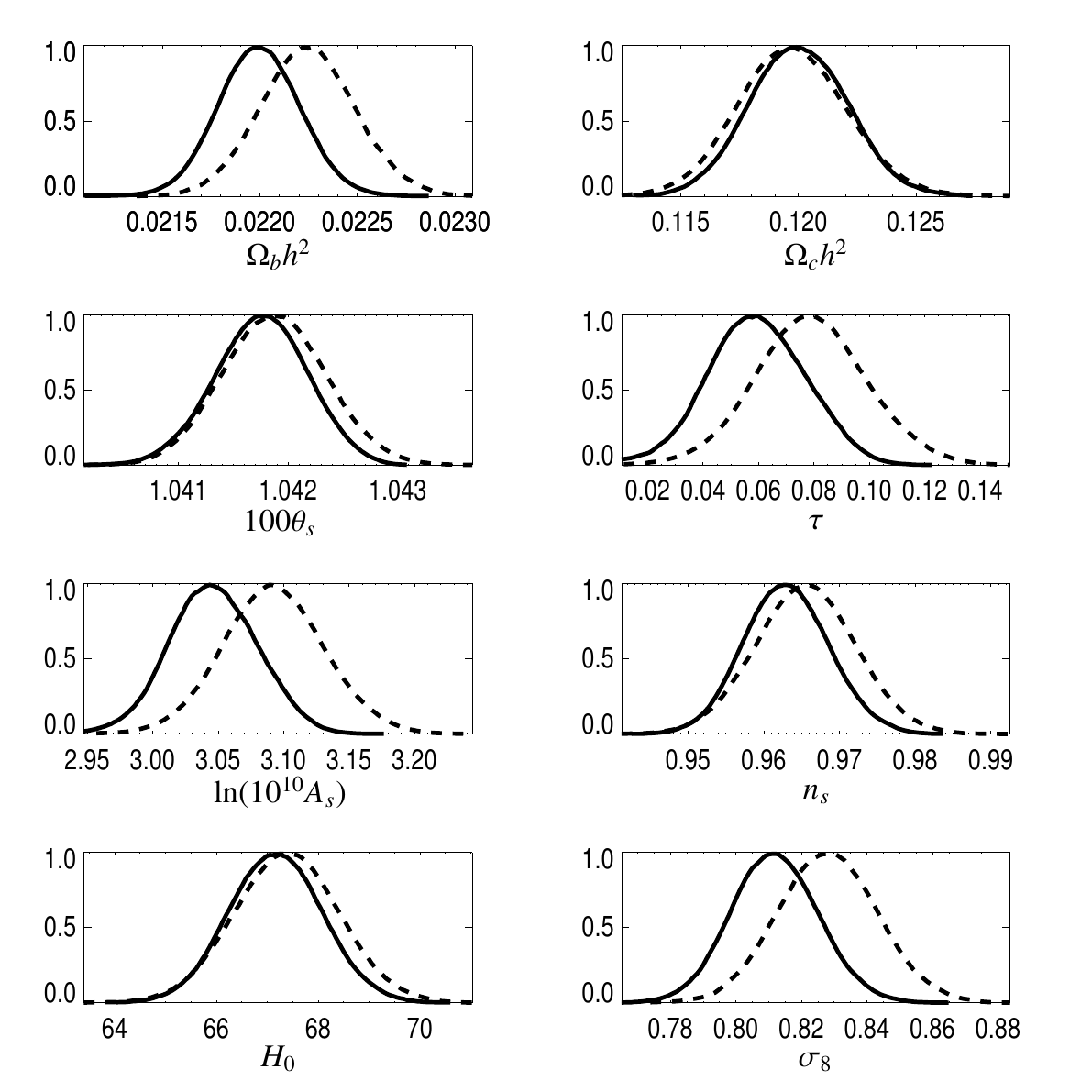}
\caption{Posterior distributions of the \lcdm\ cosmological parameters
  obtained with our regularized likelihood (\hlp+\lowTEB+VHL, full line),
  compared to the \planck\ baseline result (\plik+\lowTEB, dashed
  line). We recall that this former also includes some VHL information 
  through an SZ correlation discussed in \params.
  The last row shows derived parameters.}
\label{fig:lcdmcompare}
\end{figure}

We have shown how the \hlp\ likelihood is regularized by including the \vhl\ data.
We have checked that it leads to results that are fully compatible with the \bflike\ likelihood for $\tau$ and that their combination leads to an $\Alens$ value that is now compatible with one.
We then combine the three likelihoods and fix $\Alens$ to one, to evaluate the impact on \lcdm\ parameters.
The comparison with the \planck\ published
result is shown in Fig.~\ref{fig:lcdmcompare}. We note that:
\begin{itemize}
\item  \plik\ and \hlp\ likelihoods essentially share  the same data;
\item the problem pointed out by $\Alens$ is not a second-order
  effect: it directly affects $\Omb$, $\tau$, and $\As$ results;
\item our regularized likelihood provides a lower $\sigma_8$ estimate.
\end{itemize}

These results are obtained using only CMB data.
To check the overall consistency and constrain the parameters,
we also make use of some robust extra-information by including recent Baryon Acoustic Oscillations (BAO) and
supernovae (SN) results.

The BAO are generated by acoustic waves in the primordial fluid,
and can be measured today through the study of the correlation functions of galaxy surveys.
Owing to the fact that their measurement is sensitive to different systematic errors than the CMB, they
help break the degeneracies, and are therefore further used in this paper to constrain
the cosmological parameters.  Here, we have used:
the acoustic-scale distance ratio $D_V(z)/r_{drag}$
measurements\footnote{$D_V(z)$ is a function of the redshift ($z$) and
can be expressed in terms of the angular diameter
distance and the Hubble parameter, $r_{drag}$, is the comoving sound
horizon at the end of the baryon drag epoch.} from the 6dF Galaxy Survey at $z=0.1$~\citep{Beutler:2014yhv},
and from BOSS-LowZ at $z=0.32$. They have been combined with the BOSS-CMASS anisotropic 
measurements at $z=0.57$, considering both the line of sight and the
transverse direction, as described in~\citet{BAO_Anderson2014}. 

Type Ia supernovae had a major role in the discovery of late
time acceleration of the Universe and constitutes a powerful
cosmological probe complementary to CMB constraints.
We have used the JLA compilation \citep{Betoule_2014}, which covers a
wide redshift range (from 0.01 to 1.2). 
 
 \begin{table*}[h!]     
 \begingroup
 \openup 5pt
 \newdimen\tblskip \tblskip=5pt
 \nointerlineskip
 \vskip -3mm
 \footnotesize
 \setbox\tablebox=\vbox{
     \newdimen\digitwidth
     \setbox0=\hbox{\rm 0}
     \digitwidth=\wd0
     \catcode`*=\active
     \def*{\kern\digitwidth}
     \newdimen\signwidth
     \setbox0=\hbox{+}
     \signwidth=\wd0
     \catcode`!=\active
     \def!{\kern\signwidth}
\halign{
\hbox to 0.9in{$#$\leaderfil}\tabskip=10pt&
\hfil$#$\hfil\tabskip=15pt&
\hfil$#$\hfil&
\hfil$#$\hfil \cr
\noalign{\doubleline}
\omit Parameter\hfil& $\hlp+\lowTEB$& $\hlp+\lowTEB$& $\hlp+\lowTEB$ \cr  
\omit & & \hfil $+VHL$ \hfil& $+VHL+BAO+SN$  \cr  
\noalign{\vskip 3pt\hrule\vskip 5pt}
\Omb h^2                        &***0.02220\pm0.00023   &***0.02203\pm0.00020   &***0.02211 \pm 0.00018     \cr
\Omc h^2                        &*0.1193\pm0.0022       &*0.1196\pm0.0020       &*0.1183 \pm 0.0012      \cr
100\theta_{\rm s}               &***1.04179\pm0.00043   &***1.04181\pm0.00042   &*** 1.04190 \pm 0.00038\cr
\tau                            &0.071\pm0.019          &0.058\pm0.018          &0.062 \pm 0.017       \cr
\ns                             &**0.9644\pm0.0069      &**0.9626\pm0.0055      &**0.9654 \pm 0.0040      \cr
\lnAs                           &3.068\pm0.037          &3.044\pm0.034          &3.048 \pm 0.033       \cr
\noalign{\vskip 3pt\hrule\vskip 5pt}
H_0                             &67.48\pm0.98           &67.21\pm0.91           &67.77\pm0.57           \cr
\sigma_8                        &*0.816\pm0.015         &*0.809\pm0.013         &*0.807\pm0.013         \cr
\noalign{\vskip 5pt\hrule\vskip 3pt}}}
\endPlancktablewide                
\endgroup
\caption{%
Estimates of cosmological parameters using MCMC techniques for the six \lcdm\ parameters.
First with our likelihood (\hlp) and \bflike. Then with our regularized likelihood (\hlp+\lowTEB+VHL, second column) and further
adding some BAO and SN data (third column). Here $\theta_s$, as computed by \CLASS, represents the exact angular size of the sound horizon and should not be identified with the \COSMOMC\ $\theta_{MC}$ parameter (see Appendix~\ref{app:minuit}).
}                          
\label{Tab::LCDMHLP} 

\end{table*}                        

Table~\ref{Tab::LCDMHLP} gives the results obtained with
\hlp+\bflike\ (i.e., using only \planck\ data), then adding ACT+SPT likelihoods
 (i.e., only CMB) and finally also adding  the BAO and SNIa likelihoods.
Using only \planck\ data, the \hlp+\bflike\ results (first column in Table \ref{Tab::LCDMHLP})
are almost identical to the ones reported for \q{$\planckTTonly\dataplus\lowP$} in \params, but for $\tau$ and $\As$ that are smaller, as explained in Sects.~\ref{sec:hlp_results} and~\ref{sec:plik_tau}. 

As discussed throughout this paper, adding the \vhl\ data (second column) releases the tension on the optical depth, leading to a value of $\tau$ around $0.06$, as can be anticipated from Fig. \ref{fig:hlp_vhl_tauprof}.
We also see a slight shift of $\Omb h^2$ that is difficult to analyze since, 
at this level of precision, this parameter enters several areas of the Boltzmann
computations \citeg{huwhite97}.
Then, adding the BAO and SN data (third column) increases the precision on the parameters but does not change substantially their value.

We note that in all cases, $\sigma_8$ is stable, e.g., \hlp+\bflike\ gives $\sigma_8=0.816\pm0.015$.
This is only in mild tension with other astrophysical determinations such as weak lensing~\citep{Heymans:2013fya} and Sunyaev-Zeldovich cluster number counts~\citep{planck2014-a30}.


\section{Conclusion}
In this work, we have investigated the deviation
of the $\Alens$ parameter from unity and have found it to be of $2.6\sigma$, using the \plik\
high-$\ell$ and \bflike\ low-$\ell$ \planck\ likelihoods.
For these demanding tests, we chose to consistently use the
profile-likelihood method, which is well-suited to such studies.
We first showed how this $\Alens$ deviation is related to
a difference in the $\tau$ estimations when performed using \plik \
or \bflike\ alone.  

The \hlp\ likelihood has been built based on different foreground and
nuisance parametrization.
We have shown that \hlp\ alone  only very loosely  constrains
$\Alens$  towards high values, and that its $\tau$ estimate lies
closer to the low-$\ell$ measurement, albeit on the high-end side.  

We then added to \hlp, high angular resolution CMB data from the
ground-based ACT and SPT experiments to
further constrain the high-$\ell$ part of the CMB power spectrum and the foreground
parameters. Cosmological parameters derived from this setup are shown
to be more self-consistent, in particular the reconstructed $\tau$
value is  coherent with the low-$\ell$ determination that was extracted from
the  \bflike\ likelihood. They also pass the $\Alens = 1.0$ test. 

We have shown that this regularization is quite robust against the
details of the \vhl\ datasets used, and specific foreground
hypotheses. We have also shown that it is not only related to a better
determination of the foreground amplitudes but also seems to lie in
their correlations.

The cosmological parameters determined  from this combined CMB
likelihood are also stable when adding BAO and SNIa likelihoods. 
This is, in particular, the case for $\sigma_8$ that we always find close to 0.81.
With respect to the cosmological parameters derived by the \planck\
collaboration, the main differences concern $\tau$ and $\As$, to which
the former is directly correlated, and $\Omega_b h^2$, which shifts by
a fraction of $\sigma$.  
Other parameters are almost identical.

Improving on $\Alens$ is a delicate task and
it seems that the source of the regularization cannot be easily pinpointed.
The choices of the \hlp\ likelihood impact on the correlations between
all the parameters, yielding a $\tau$ estimate in smaller tension with the
\lowell\ likelihood. But this was not sufficient to
relieve the $\Alens$ tension. It is only by further constraining
foregrounds using \vhl\ likelihoods that we were able to obtain a
coherent picture over a broad range of multipoles. 

One cannot exclude that the $\Alens$ deviation from unity
still partly  results from an incomplete  accounting of
some residual systematics in the ACT, SPT, or \planck\ data. 

During the review of this article, the \planck\ Collaboration released
an estimate of the optical depth based on HFI cleaned maps
\citep{planck2014-a25} and using the \texttt{Lollipop}
likelihood~\citep{lollipop}. 
This new \lowell-only result, $\tau=0.058\pm0.012$, 
increases the tension with the \hiell\ likelihoods to $2.6\sigma$
with \plik\ and $1.5\sigma$ with \hlp\ and is fully compatible with our
\hlp+VHL combination (\refeq{tauhlpvhl}).

\begin{acknowledgements}
We thank Guilaine Lagache for building and providing the point-source
masks cleaned from Galactic compact structures,
Marian Douspis for providing the SZ templates,
and Marc Betoule for the development of the {\tt C} version of
the {\tt JLA} likelihood.
\end{acknowledgements}

\appendix
\section{\CLASS\ vs. \CAMB\ results}
\label{app:classvscamb}

Our results were obtained using the \CLASS\ \texttt{v2.3.2} Boltzmann code,
 which computes the temperature and polarization power spectra by evolving the cosmological background and
 perturbation equations.
Besides being very clearly written in \texttt{C} and modular, 
\CLASS\ has a very complete set of precision parameters in  a
single place that enables us to study them efficiently. This makes it a perfect tool for developing modern
high-precision cosmological projects.

We revisit the agreement between \CLASS\ and \CAMB\ at the level
required to estimate $\Alens$. 
For this purpose, we fix the cosmology to the \planckTT\ best fit and run \CLASS\ with three different
settings:
\begin{itemize}
\item the \CLASS\ default ones;
\item the high-quality (HQ) ones that are being used to obtain some
smooth profile-likelihoods and are given in Table~\ref{tab:class_high-prec};
\item the very high-quality (VHQ) ones corresponding to the maximal
precision one can reasonably reach on a cluster\footnote{This single shot run last
several minutes on 8 cores and requires about 30GB of memory} and that are provided in \CLASS\
in the file \verb=cl_ref.pre=.
\end{itemize}

\begin{figure}[!ht]
\centering
\includegraphics[width=88mm]{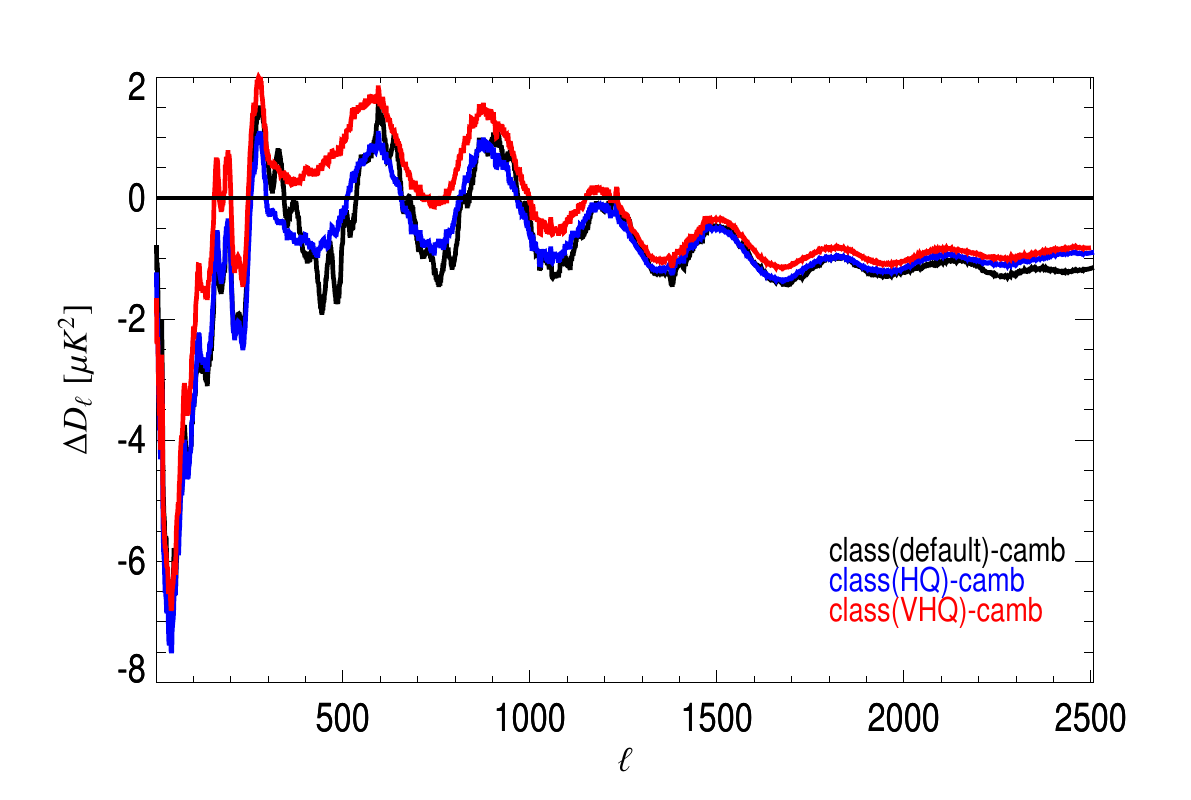}
\caption{Difference of
  $D_\ell\equiv \ell(\ell+1)\clTT/2\pi$ power
  spectra between \CLASS\ and \CAMB\ for the same fiducial cosmology. 
  For \CLASS\ several precision settings are tested:
  the default ones ( in black),
  the high-quality (in blue) and very high-quality ones (in red) as described in the text.}
\label{fig:compspectra}
\end{figure}

For \CAMB,\ we use the corresponding spectrum released in the \planck\
Legacy Archive.
We compare the spectra on Fig.~\ref{fig:compspectra}
Generally speaking the agreement is at the \muk\
level, which is below one percent agreement over the whole range. In practice, this means
it does not impact on the \lcdm\ parameters, as discussed in \liket.

The \lowell\ difference is not very important
for data analysis since experiments are limited there by the cosmic variance.
The asymptotic $\simeq 1\muk$ offset is more worrying and some \lcdm\ extensions could be sensitive 
to this difference.
Since the \CLASS\ spectrum is slightly lower than the \CAMB\ one precisely
in the \hiell\ region where $\Alens$ has the largest impact, it should lead to a higher $\Alens$ value.
This is indeed observed and discussed in Sect.~\ref{sec:alens}, but
has a small effect on $\Alens$.

Finally we note that all \CLASS\ settings give similar results, 
in particular in their \hiell\ region, and that our HQ settings (used
for profile-likelihoods) are
smoother than the default ones and similar to the most extreme ones (VHQ).

\section{Optimization of profile-likelihoods}
\label{app:minuit}

The computation of genuine profile-likelihoods requires exquisite
minimization precision. Since the 68\% confidence intervals are obtained by thresholding
the profiles at 1, we need a precision on $\chi^2_\text{min}$, well below 0.1,
which is challenging for a numerical method without analytic gradients. 
It has already been shown in \pip\ how using the \texttt{Minuit} package and
increasing the \CLASS\ precision parameters to high values (still
keeping a reasonable computation time) enables us to achieve this goal.

\begin{table}[h!]
\begin{center}
\begingroup
\openup 5pt
\newdimen\tblskip \tblskip=5pt
\nointerlineskip
\vskip -3mm
\footnotesize
\setbox\tablebox=\vbox{
    \newdimen\digitwidth
    \setbox0=\hbox{\rm 0}
    \digitwidth=\wd0
    \catcode`"=\active
    \def"{\kern\digitwidth}
    \newdimen\signwidth
    \setbox0=\hbox{+}
    \signwidth=\wd0
    \catcode`!=\active
    \def!{\kern\signwidth}
\halign{
#\hfil\tabskip=1.5em&\hfil$#$\hfil\tabskip=0pt\cr
\noalign{\doubleline}
\omit\hfil \CLASS\ parameter\hfil&\omit\hfil Value\hfil\cr
\noalign{\vskip 3pt\hrule\vskip 5pt}
tol\_background\_integration&10^{-3}\cr
tol\_thermo\_integration&10^{-3}\cr
tol\_perturb\_integration&10^{-6}\cr
reionization\_optical\_depth\_tol&10^{-5}\cr
l\_logstep&1.08\cr
l\_linstep&25\cr
perturb\_sampling\_stepsize&0.04\cr
delta\_l\_max&1000 \cr
accurate\_lensing & 1\cr
\noalign{\vskip 5pt\hrule\vskip 3pt}
} 
} 
\endPlancktable
\endgroup
\caption{\label{tab:class_high-prec}
High-precision settings of the \CLASS\ non-default parameters used
in the profile-likelihood constructions. The last parameter
(accurate\_lensing) is only useful (and used) to model in details the
\vhl\ tail (typically above 4000) and is only used with our ACT
and SPT likelihoods.}
\end{center}
\end{table}

Besides revisiting our high-precision parameters that are shown in
Table~\ref{tab:class_high-prec}, we have  improved our strategy further by using
the following scheme: 
\begin{enumerate}
\item we use the \texttt{pico} software\footnote{\url{https://sites.google.com/a/ucdavis.edu/pico}} to pre-compute a
  best fit solution. \texttt{pico} is based on the interpolation
  between spectra trained on the \CAMB\ Boltzmann solver. It is
  very fast and the minimization converges rapidly owing to
  the smoothness between models.
\item  since \CAMB\ computes only an approximation of the angular size
  of the sound horizon ($\theta_{MC}$) while \CLASS\ computes it
  exactly ($\theta_s$) we perform the change of variables by fixing
  the previous best-fit estimates and performing a 1D fit to
  $\theta_s$ only.
\item we then have a good starting point to \CLASS\ and perform the
  minimization with a high-precision strategy (\texttt{Migrad},
  strategy=2 level).
\item optionally: in some rare cases, the estimated minimum is not accurate enough (this can be identified by checking the profile-likelihood continuity).
In this case, we use several random initialization points (typically 10)
near the previous minimum, perform the same minimization, and keep the
lowest $\chi^2_{min}$ solution.
\end{enumerate}

In all cases we have checked that the \texttt{pico} results are
similar to the final \CLASS\ ones but consider the latter to give
more precise results since \texttt{pico} implements only an approximation to the
$\Alens$ parameter and was trained on an old \CAMB\ version.


\section{\vhl\ consistency}
\label{app:fg}

\begin{figure*}
  \centering
\subfigure[{\tt ACT\_equat} $148\times 148$ GHz]{\includegraphics[width=.45\textwidth]{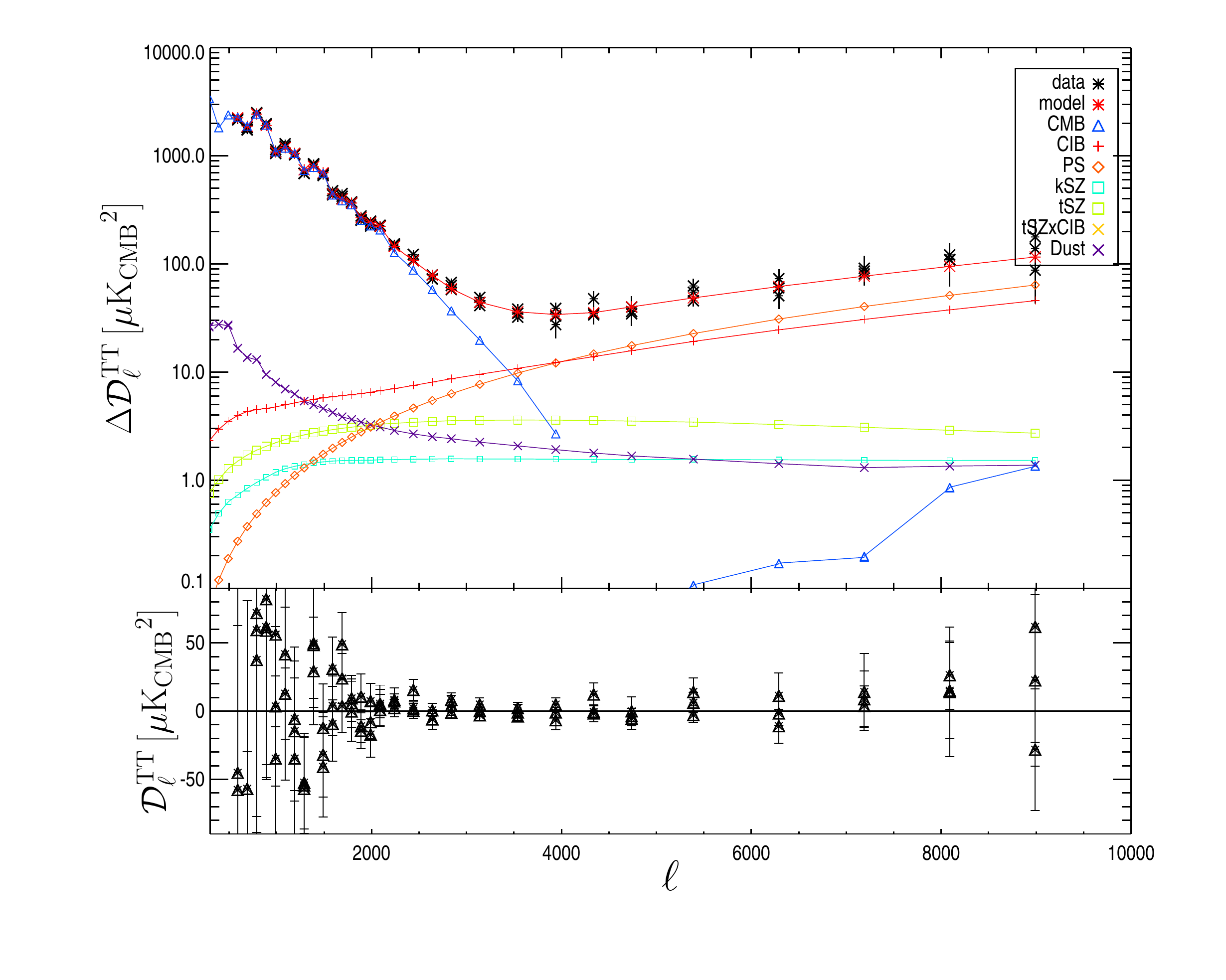}}
\subfigure[{\tt ACT\_equat} $148\times 218$ GHz]{\includegraphics[width=.45\textwidth]{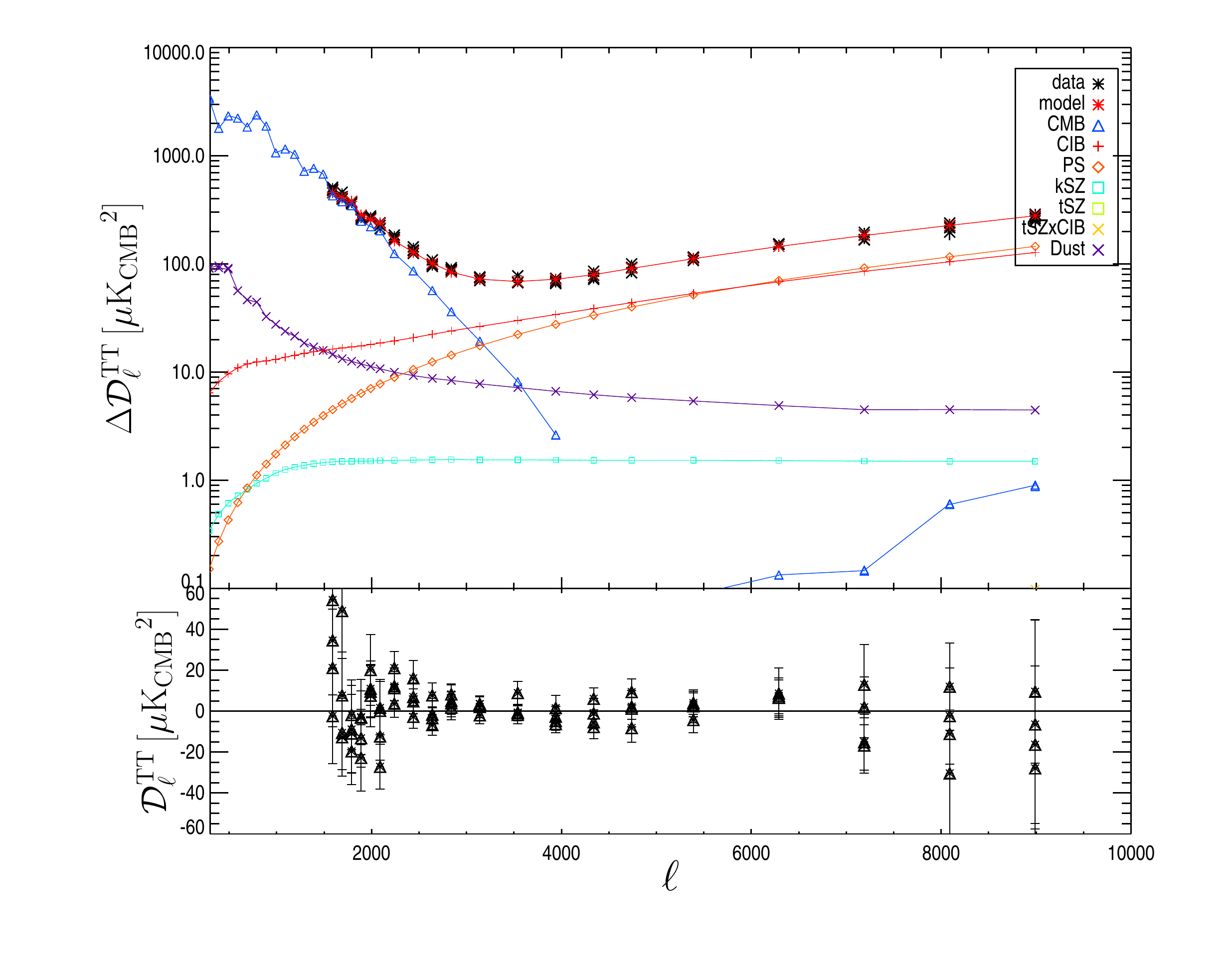}}\\
\subfigure[{\tt ACT\_equat} $218\times 218$ GHz]{\includegraphics[width=.45\textwidth]{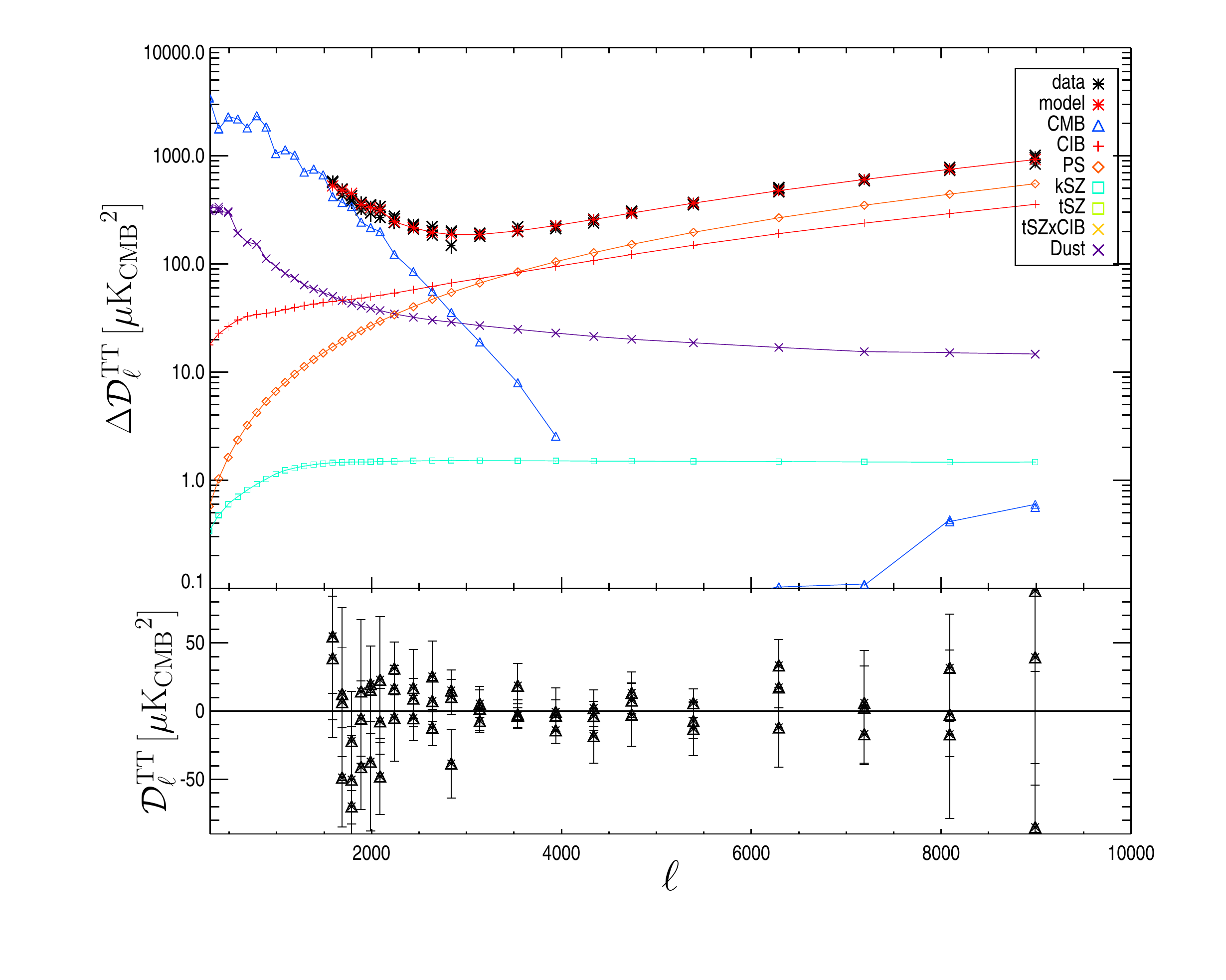}}
\subfigure[{\tt ACT\_south} $148\times 148$ GHz]{\includegraphics[width=.45\textwidth]{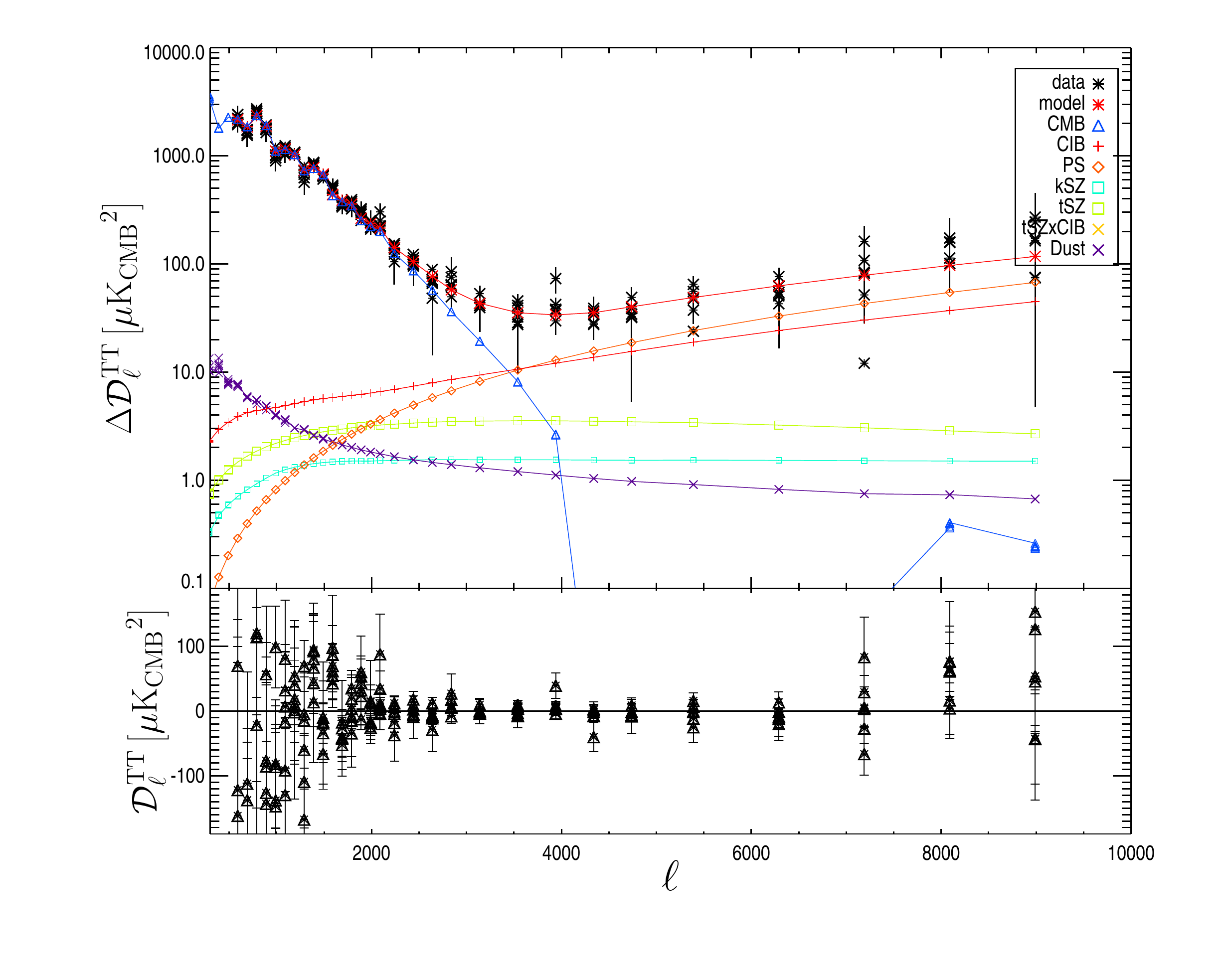}}\\
\subfigure[{\tt ACT\_south} $148\times 218$ GHz]{\includegraphics[width=.45\textwidth]{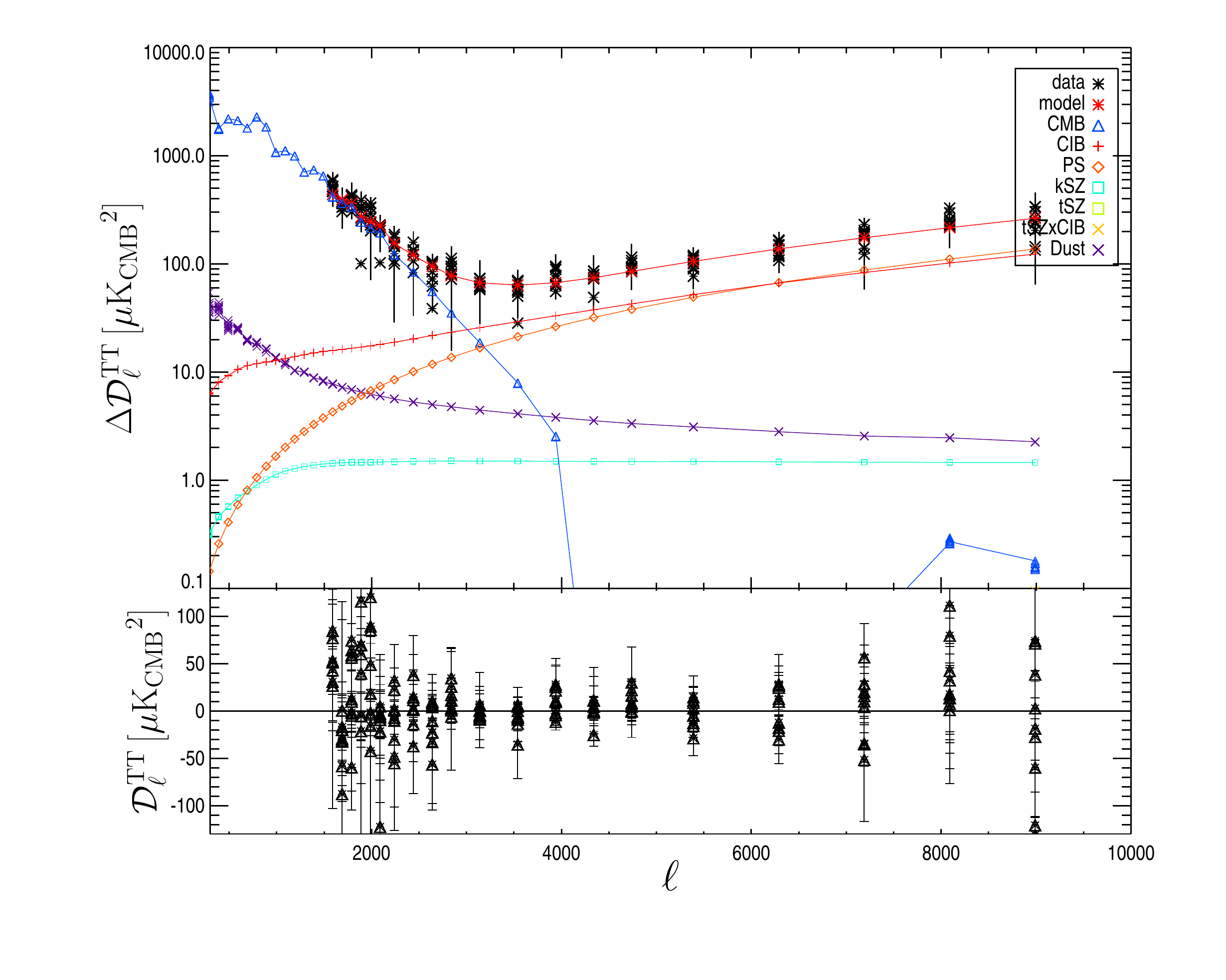}}
\subfigure[{\tt ACT\_south} $218\times 218$ GHz]{\includegraphics[width=.45\textwidth]{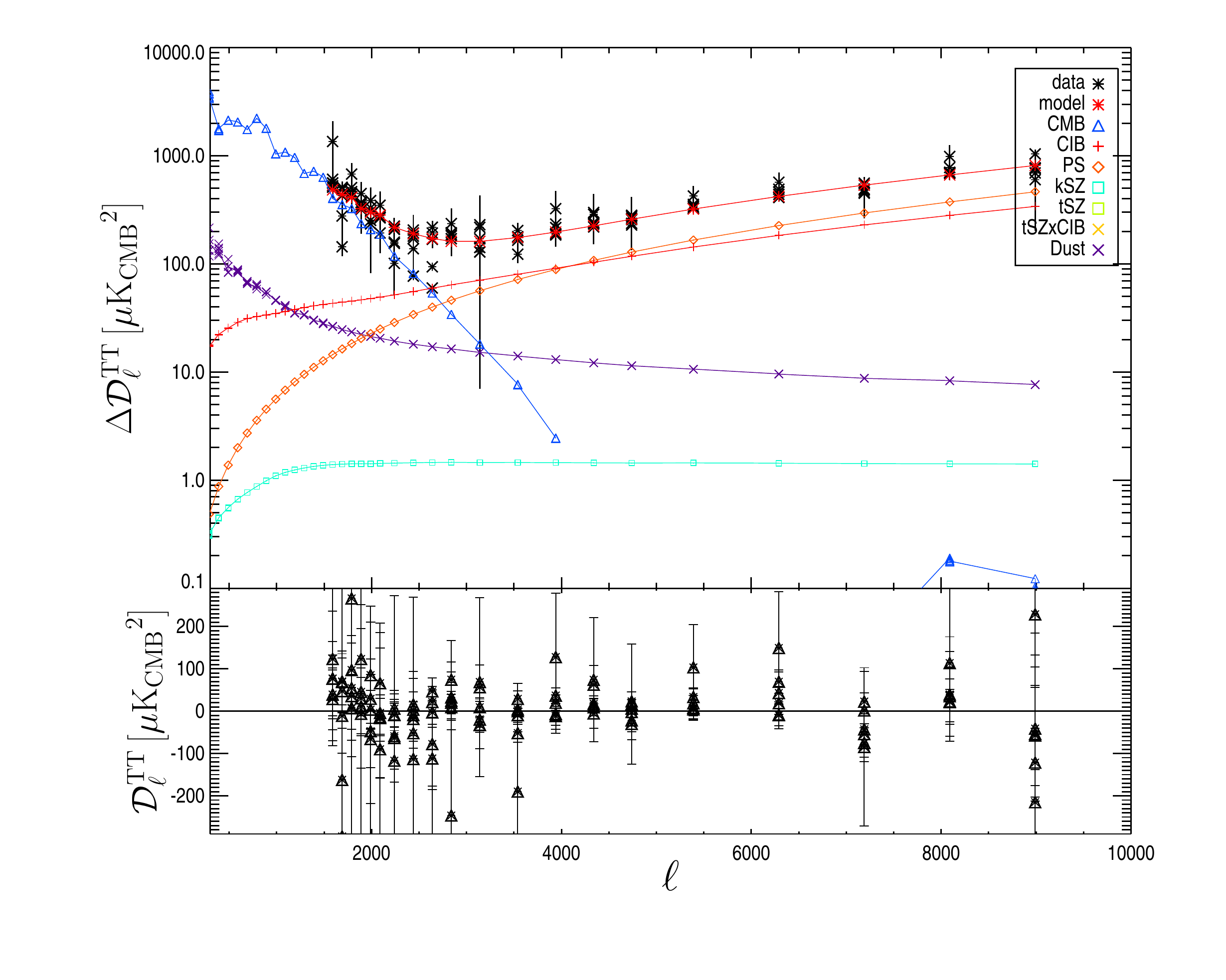}}
\caption{ \label{fig:fgdetails_ACT}Power spectra of CMB and foregrounds as fitted
by the \hlp+VHL likelihood, compared with ACT data. The use of window functions explain why the CMB component sometimes reappears at the very end of the multipole range.}
\end{figure*}

\begin{figure*}
  \centering
\subfigure[{\tt SPT\_high} $95\times 95$ GHz]{\includegraphics[width=.45\textwidth]{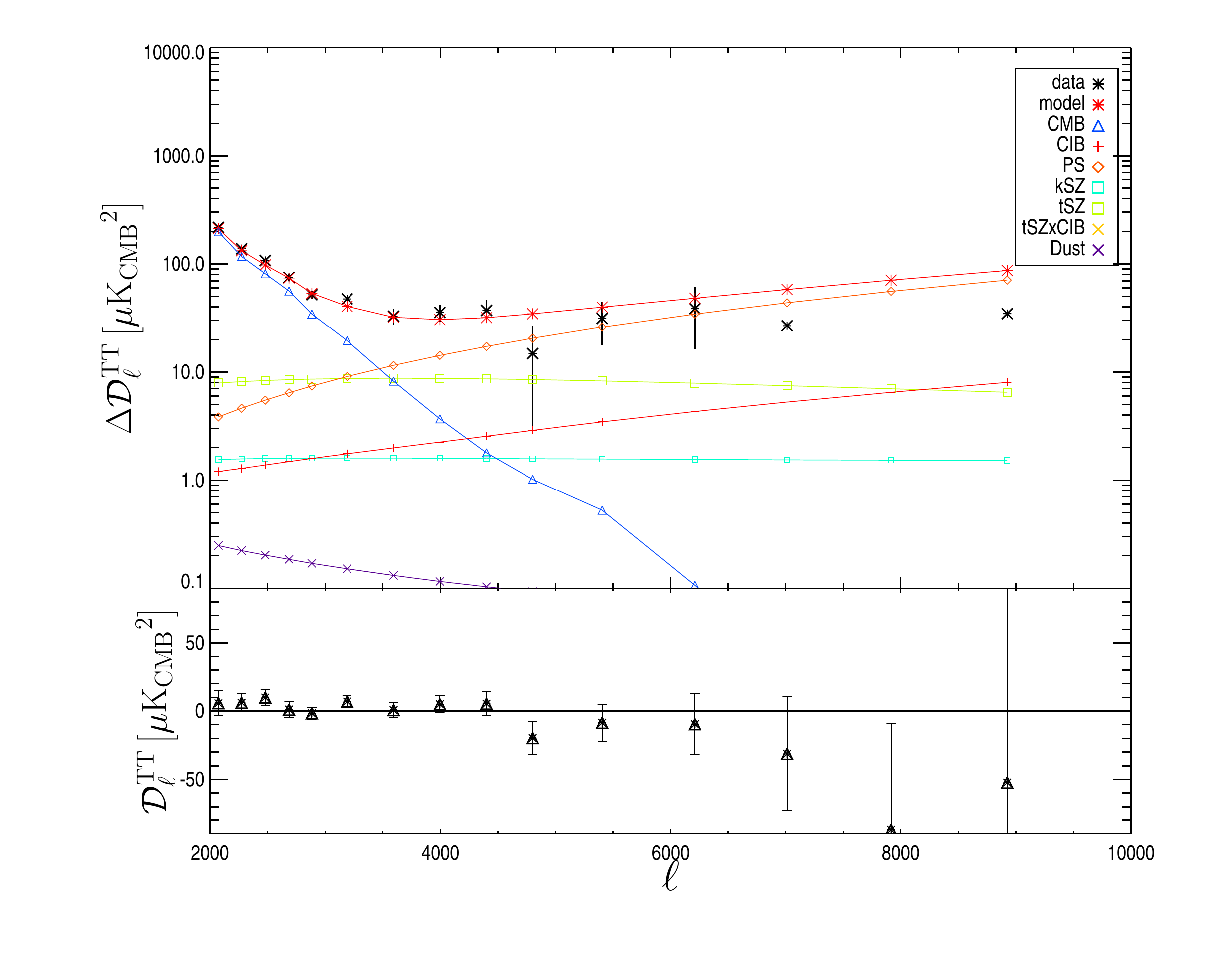}}
\subfigure[{\tt SPT\_high} $95\times 150$ GHz]{\includegraphics[width=.45\textwidth]{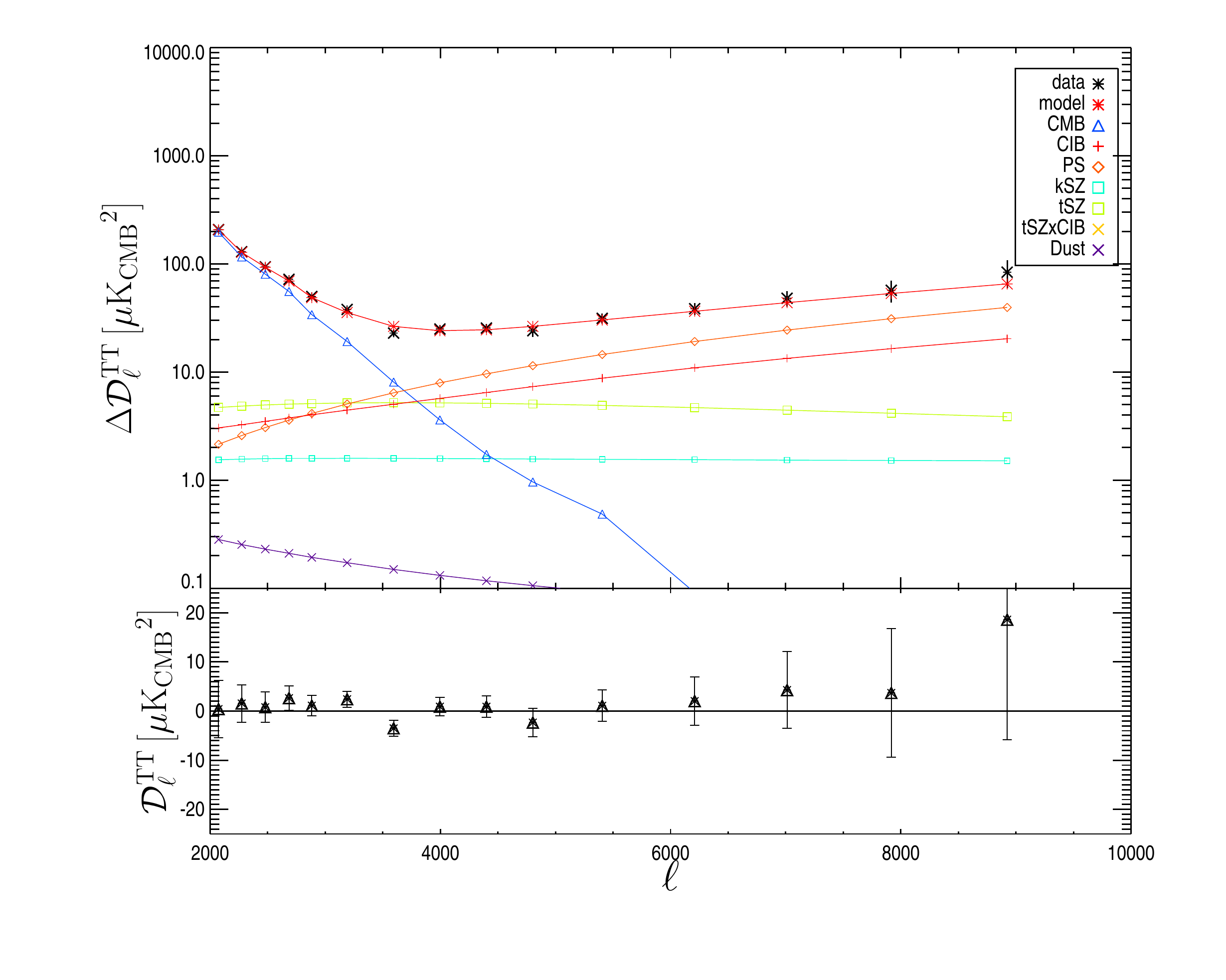}}\\
\subfigure[{\tt SPT\_high} $95\times 220$ GHz]{\includegraphics[width=.45\textwidth]{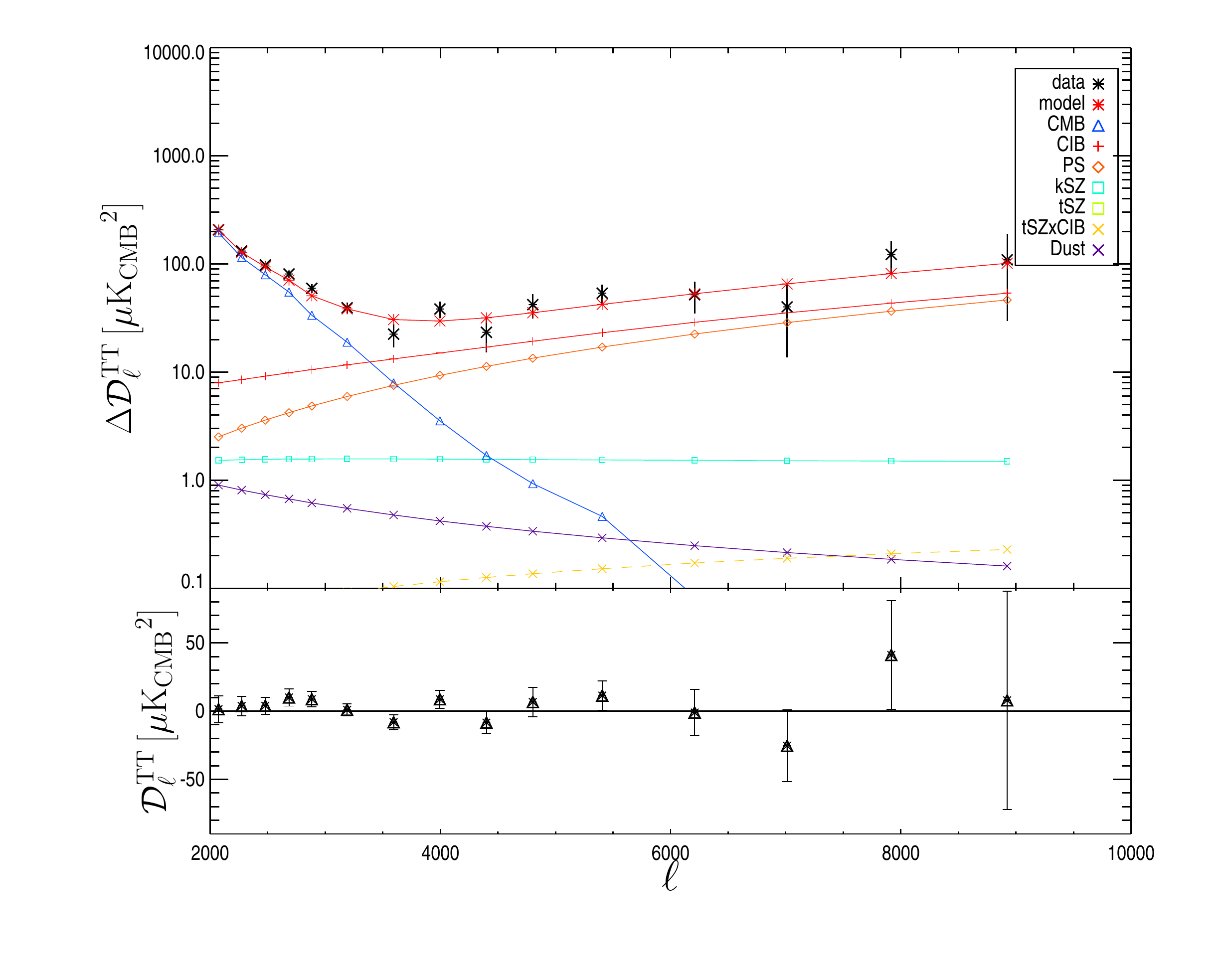}}
\subfigure[{\tt SPT\_high} $150\times 150$ GHz]{\includegraphics[width=.45\textwidth]{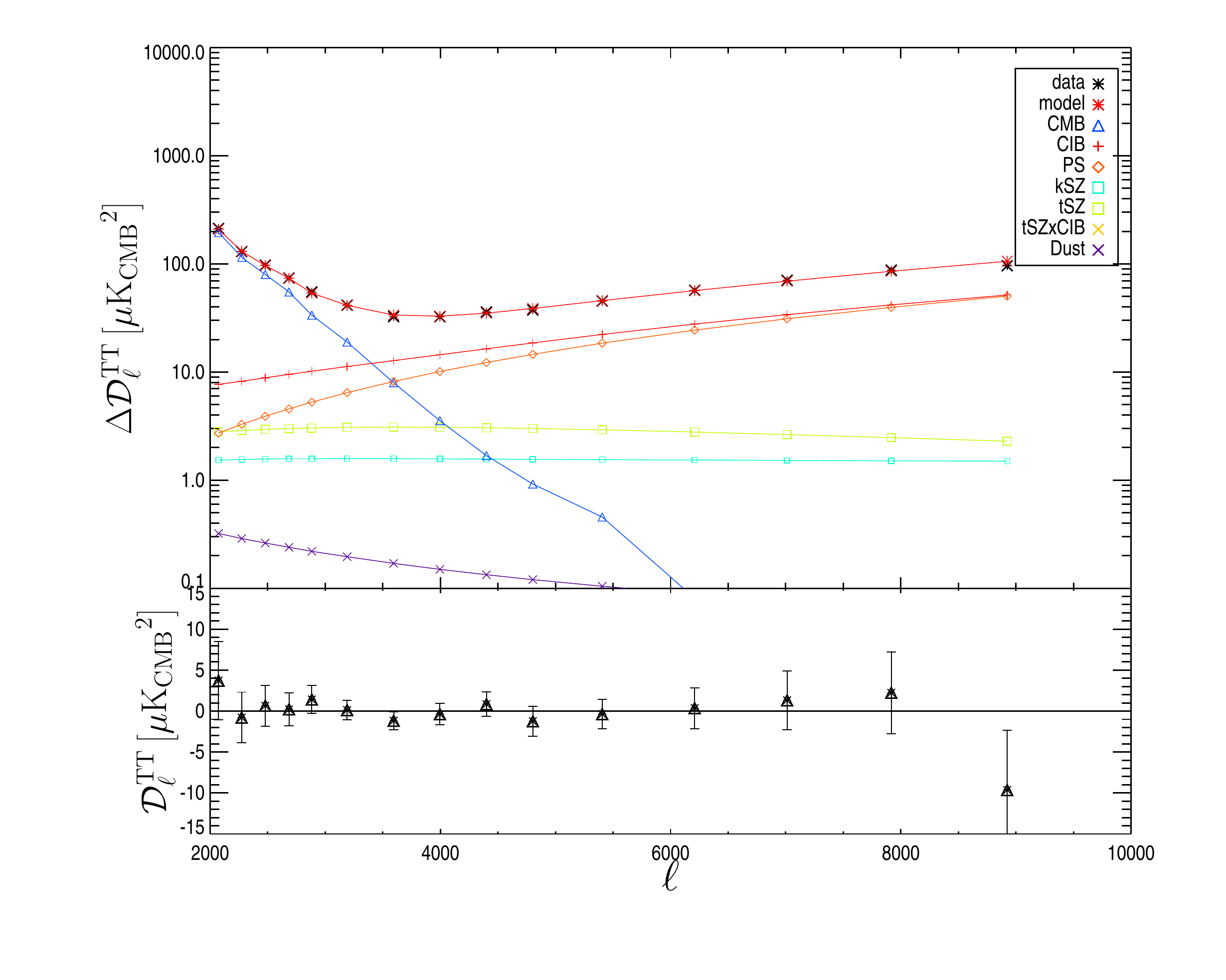}}\\
\subfigure[{\tt SPT\_high} $150\times 220$ GHz]{\includegraphics[width=.45\textwidth]{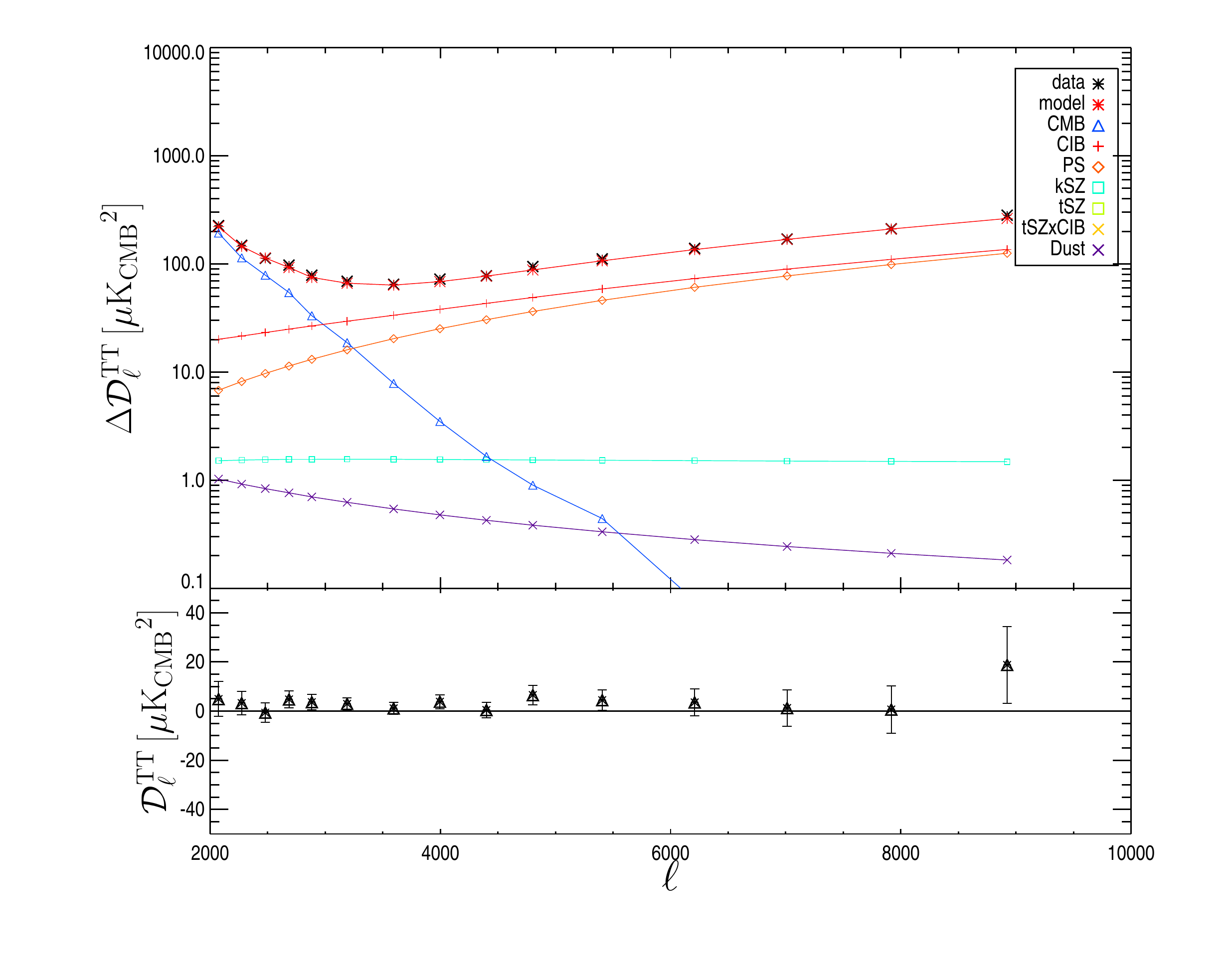}}
\subfigure[{\tt SPT\_high} $220\times 220$ GHz]{\includegraphics[width=.45\textwidth]{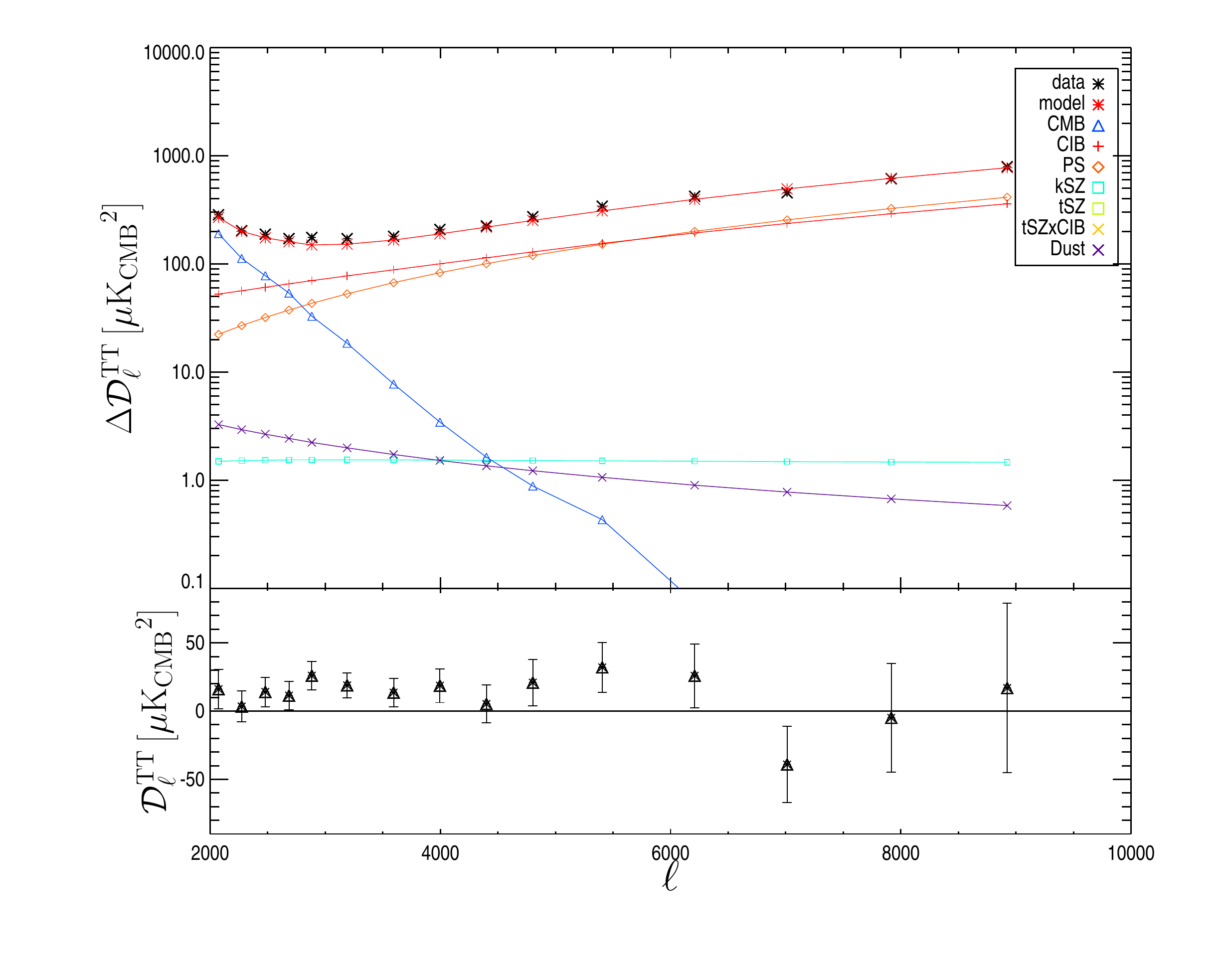}}\\
\caption{ \label{fig:fgdetails_SPT}Power spectra of CMB and foregrounds as fitted
by the \hlp+VHL likelihood compared with {\tt SPT\_high} data.}
\end{figure*}

\begin{figure*}
  \centering
\subfigure[{\tt SPT\_low} $150\times 150$ GHz]{\includegraphics[width=.45\textwidth]{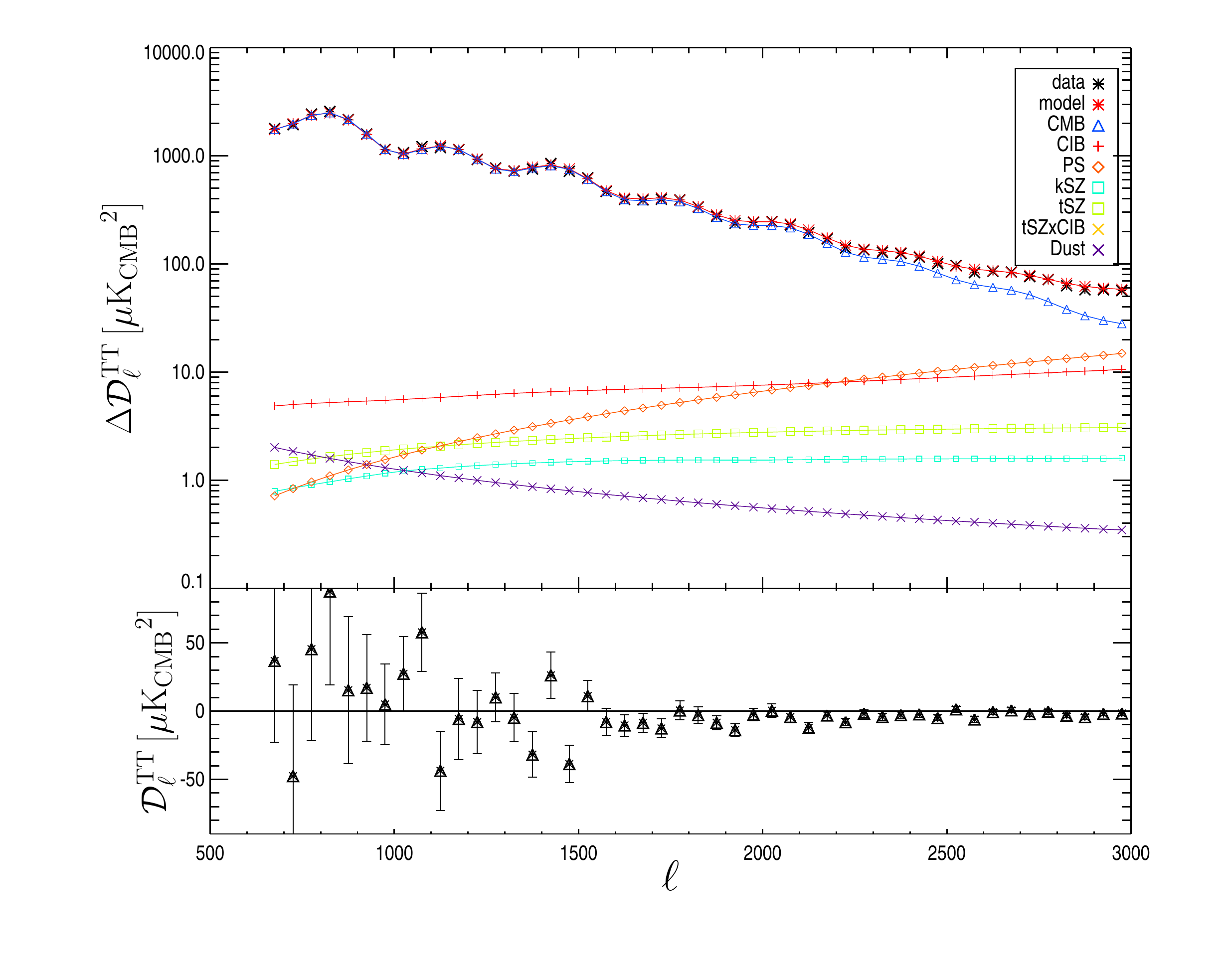}}
\caption{ \label{fig:fgdetails_SPTlow}Power spectra of CMB and foregrounds as fitted
by the \hlp+VHL likelihood compared with {\tt SPT\_low} data.}
\end{figure*}

Here we provide  more details on the checks we
performed  to assess the ACT, SPT, and \planck\ spectra compatibility in
the \vhl\ regime. 
Fig.~\ref{fig:fgdetails_ACT}, \ref{fig:fgdetails_SPT}, and \ref{fig:fgdetails_SPTlow}  show the detailed fit of each
component for each dataset. The components are obtained from the
templates scaled according to the combined \hlp+\bflike+VHL fitted cosmological and nuisance parameters.
The general agreement on the \vhl\ side is very good as shown by
the $\chi^2$ for each dataset given in
Sect.~\ref{ssect:firstvhlres}.
We note that the points are largely correlated and 
that the exact $\chi^2$ computation involves the
full covariance matrices provided by each experiment, which all contain
non-negligible off-diagonal terms.
For example, the $\chi^2$ from the {\tt SPT\_low} part is 58 for 47
degrees of freedom, which has a 13\% PTE.
Some small excess in data seems to show up in the {\tt SPT\_high} 220$\times$220~GHz 
spectrum above $\ell=3000$. In Sect.~\ref{sec:syst} we checked that this does not influence any of our results.

To check that the overlap in the data used on the 
{\tt SPT\_low}and {\tt SPT\_high} datasets, we did three additional profile likelihood analyses for $\Alens$. 
First, we ignored the 150 GHz data from {\tt SPT\_high}, recomputing accordingly the inverse covariance matrix used in the 
$\chi^2$ computation. We also tested the impact of removing in the {\tt SPT\_low} spectrum all bins above $\ell=3000$ 
(which are accounted for in the {\tt SPT\_high} part. To do this, we again recomputed the inverse covariance matrix. 
Finally, we omit in the {\tt SPT\_high} part bins at 150 GHz at $\ell <3000$. These results are reported in Table~\ref{tab:Alens_vhls_ann}. All are compatible with each other and with unity. 

 \begin{table}[h!]     

 \begingroup
 \openup 5pt
 \newdimen\tblskip \tblskip=5pt
 \nointerlineskip
 \vskip -3mm
 \footnotesize
 \setbox\tablebox=\vbox{
     \newdimen\digitwidth
     \setbox0=\hbox{\rm 0}
     \digitwidth=\wd0
     \catcode`*=\active
     \def*{\kern\digitwidth}
     \newdimen\signwidth
     \setbox0=\hbox{+}
     \signwidth=\wd0
     \catcode`!=\active
     \def!{\kern\signwidth}
 \halign{
 \hbox to 1.8in{$#$\leaderfil}\tabskip=10pt& \hfil$#$\hfil \cr
 \noalign{\doubleline}
 \omit \hfil Dataset \hfil&\omit\hfil $\Alens$ \hfil \cr
 {\tt SPT\_low}+{\tt SPT\_high}^{a}+{\tt ACT} & 1.00 \pm 0.08 \cr
 {\tt SPT\_low}^b+{\tt SPT\_high}+{\tt ACT} & 1.05\pm0.08\cr
 {\tt SPT\_low}+{\tt SPT\_high}^c+{\tt ACT} & 0.99\pm0.08\cr
 \noalign{\vskip 5pt\hrule\vskip 3pt}
 } 
 } 
 \endPlancktable
 \endgroup
 \caption{%
Results on $\Alens$ using \hlp+\lowTEB\ and VHL, making various attempts at removing the correlated part 
between {\tt SPT\_high} and {\tt SPT\_low}:  (a) w/o 150 GHz data ; (b) w/o overlapping bins; (c) w/o overlapping bins.
\label{tab:Alens_vhls_ann} }
\end{table}

\pagebreak
\bibliography{refs}{}
\bibliographystyle{aa_arxiv}

\raggedright
\end{document}